%% file: manuscript_arxiv.tex
\documentclass[a4paper,fleqn,usenatbib]{mnras}

\usepackage[T1]{fontenc}
\usepackage{ae,aecompl}

\usepackage{graphicx}	
\usepackage{amsmath}	
\usepackage{amssymb}	

\newcommand{\msun}{\ensuremath{{M}_\odot}}

\newcommand{\swift}{\textit{Swift}}

\newcommand{\boldchange}[1]{#1}

\title[SUMaC I]{Swift Ultraviolet Survey of the Magellanic Clouds (SUMaC). I.
Shape of the Ultraviolet Dust Extinction Law and Recent Star Formation History of the Small Magellanic Cloud}

\author[L.~M.~Z. Hagen et al.]{
Lea M.~Z.\ Hagen,$^{1,2}$\thanks{E-mail: lea.zernow.hagen@gmail.com}
Michael H.\ Siegel,$^{1}$
Erik A.\ Hoversten,$^{3}$
Caryl Gronwall,$^{1,2}$
\newauthor
Stefan Immler,$^{4}$
and
Alex Hagen$^{1,2}$
\\
$^{1}$Department of Astronomy and Astrophysics, The 
Pennsylvania State University, University Park, PA 16802, USA \\
$^{2}$Institute for Gravitation and the Cosmos, The 
Pennsylvania State University, University Park, PA 16802, USA \\
$^{3}$4363 Kaufmanis Way, Eagan, MN 55123, USA
$^{4}$NASA Goddard Space Flight Center, Greenbelt, MD 20771, USA
}

\date{Accepted XXX. Received YYY; in original form ZZZ}

\pubyear{2016}

\begin{document}
\label{firstpage}
\pagerange{\pageref{firstpage}--\pageref{lastpage}}
\maketitle

\begin{abstract}

We present the first results from the Swift Ultraviolet Survey of the Magellanic Clouds (SUMaC), the \boldchange{highest resolution ultraviolet (UV) survey} of the Magellanic Clouds yet completed.  In this paper, we focus on the Small Magellanic Cloud (SMC).  When combined with \boldchange{multi-wavelength} optical and infrared observations, the three near-UV filters on the Swift Ultraviolet/Optical Telescope are conducive to measuring the shape of the dust extinction curve and the strength of the 2175~\AA\ dust bump.  We divide the SMC into UV-detected star-forming regions and large $200''$ (58~pc) pixels and \boldchange{then} model the spectral energy distributions using a Markov Chain Monte Carlo method to \boldchange{constrain} the ages, masses, and dust curve properties.  We find that the majority of the SMC has a 2175~\AA\ dust bump, which is larger to the northeast and smaller to the southwest, and that the extinction curve is universally steeper than the Galactic curve.  We also derive a star formation history and find evidence for peaks in the star formation rate at 6-10~Myr, 30-80~Myr, and 400~Myr, the latter two of which are consistent with previous work.

\end{abstract}


%
%



\defcitealias{gordon03}{G03}
\defcitealias{zaritsky02}{Z02}

\section{Introduction}

\begin{figure*}
	\centering
	\includegraphics[trim = 5mm 105mm 15mm 35mm, clip=true, scale=0.5]{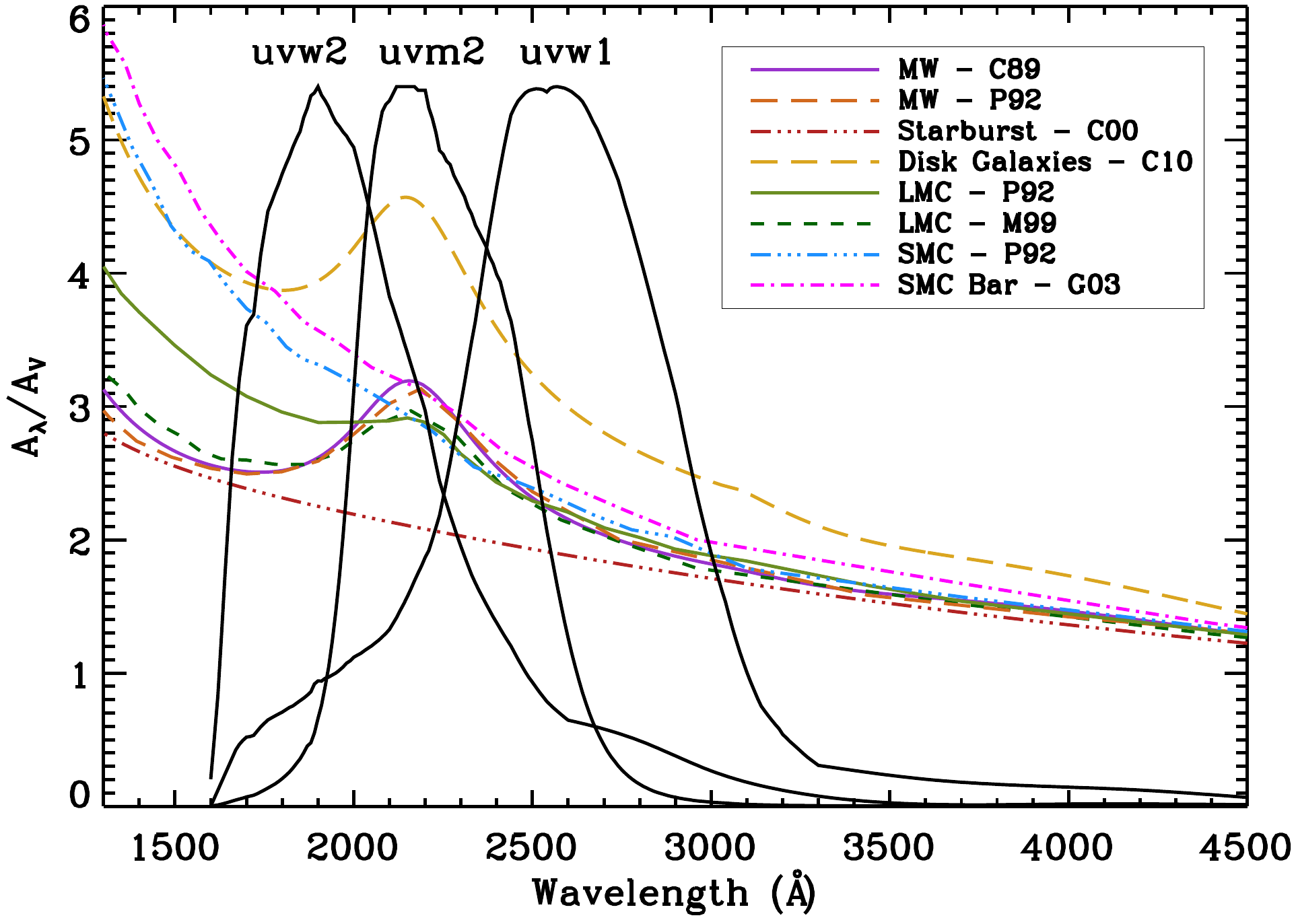}
	\includegraphics[trim = 5mm 105mm 60mm 35mm, clip=true, scale=0.5]{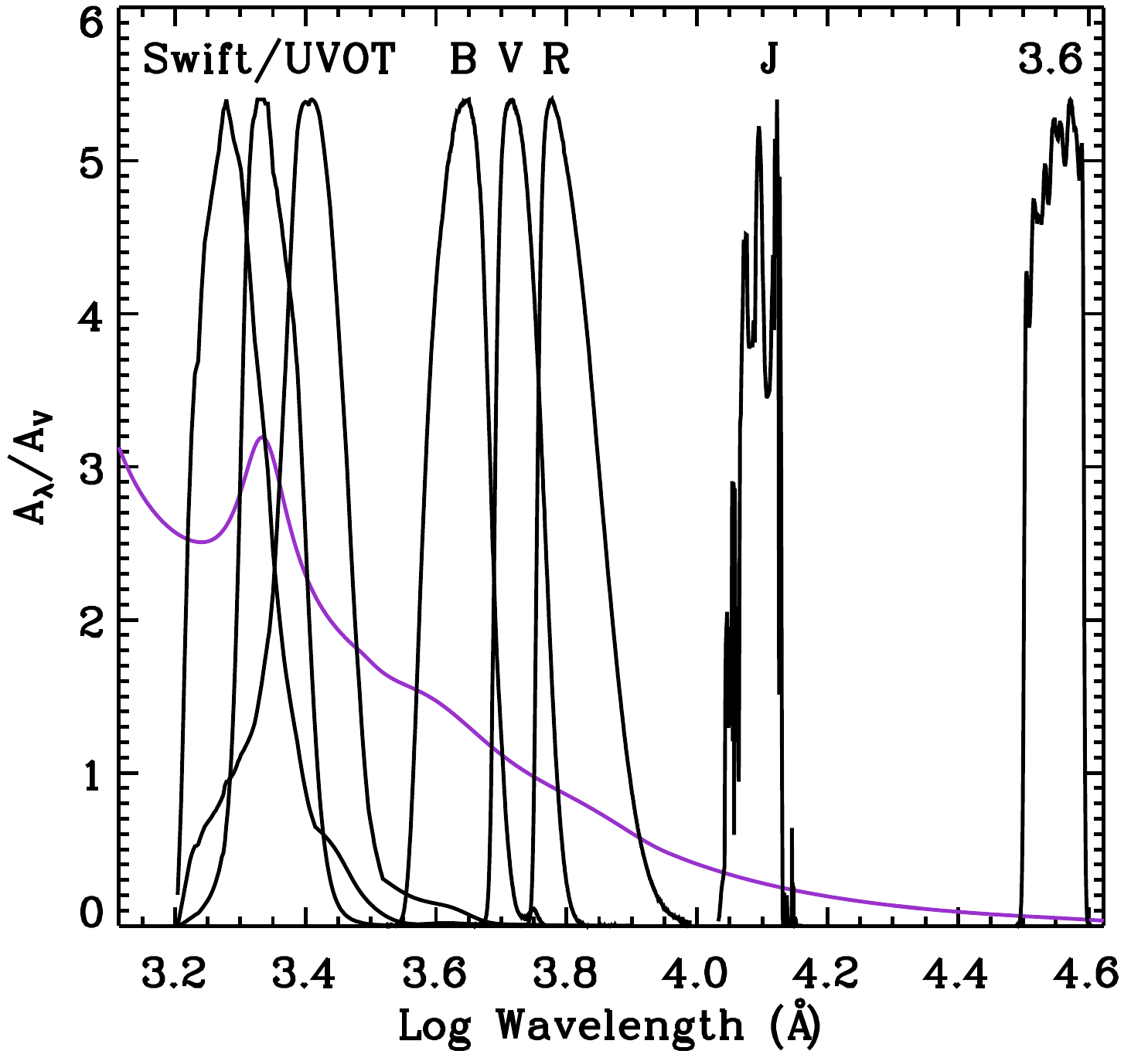}
	\caption{\textit{Left}: Several common dust extinction curves. Note the extinction curves' differing slopes and the presence (or lack) of the 2175~\AA\ dust bump.  Overlaid are the normalized filter transmission curves for the $uvw2$, $uvm2$, and $uvw1$ filters, showing how their placement is useful for constraining the properties of the dust extinction curve.
	\textit{Right}: The passbands used in this work compared to a representative Milky Way dust extinction curve.
	\textit{References}: \citet[C89]{cardelli89}, \citet[P92]{pei92}, \citet[M99]{misselt99}, \citet[C00]{calzetti00}, \citet[G03]{gordon03}, \citet[C10]{conroy10b}}
	\label{fig-dust}
\end{figure*}

When measuring the ultraviolet (UV) emission of a galaxy, it is necessary to correct for any internal dust extinction, which has a range of systematic and statistical uncertainties.
There are many prescriptions available to make this correction.  As shown in Fig.~\ref{fig-dust}, in the optical and near-infrared, the correction is small and the various prescriptions agree within uncertainties, but in the UV, they tend to diverge.  
Some, such as those from \citet[][\citetalias{gordon03}]{gordon03} and \citet{conroy10b}, are fairly steep in the UV, whereas others \citep{cardelli89,misselt99,calzetti00} are shallower.  There is also a non-ubiquitous bump in the extinction curve at 2175~\AA, first noted in \citet{stecher65}.  All of the curves plotted in Fig.~\ref{fig-dust}, with the exception of that from \citet{conroy10b}, are derived from between 5 and 30 measurements, so there is substantial room for improvement in the variability of the dust curves.
It is worth noting that the dust curves in Fig.~\ref{fig-dust} are not of uniform origin: the dust curve can be measured along a single line of sight, such as a star, or it can be an ensemble measurement averaged over many lines of sight and dust obscurations, such as for a star cluster or galaxy.

There are many ways that one can choose to quantify the dust extinction.  One way is to measure $R_V$, which follows the slope of the curve, typically quantified as
\begin{equation}
A_V = R_V \ E(B-V) \ ,
\end{equation}
where $A_V$ is the total extinction in the $V$-band and $E(B-V)$ is the colour excess.  
This is expanded to quantify the attenuation in infrared (IR) to far-UV in \citet{cardelli89} using a series of polynomial fits, with the 2175~\AA\ bump overlaid as a Drude profile \citep{bohren83}.
Another way to quantify the curve is described in \citet{noll09}, in which variation of the parameter $\delta$ changes the power law slope and $E_b$ varies the strength of the 2175~\AA\ bump. 
A third common measurement paradigm, first used in \citet{fitzpatrick90}, uses four basic components: a term linear in $\lambda^{-1}$, a far-UV curvature term, a Drude profile, and an overall offset; because this has seven free parameters, it is best suited to modeling spectra.
In this paper, we will be utilizing the \citet{cardelli89} $R_V$ and bump strength formalism, with variable bump strength quantified in the appendix of \citet{conroy10b}.


A variety of methods have been utilized to derive dust curves, many of which are noted in Table~\ref{prev_smc}.
Traditionally, measuring the shape of the dust extinction curve has been observationally intensive.  Until recently, most measurements have used UV spectra from the International Ultraviolet Explorer \citep{boggess78}, often in combination with additional optical or near-infrared (NIR) data.  
The pair method, used in many papers, requires observing spectra of a dust-obscured star and an unobscured star of the same spectral type, then comparing the spectra.  This can also be accomplished by instead comparing to an unreddened model stellar spectrum.
One can alternatively follow the method in \citet{calzetti94}, in which the UV spectral slope and Balmer emission lines in a galaxy are used together to compare to models of dust.
Another method is to compare the colour excess between the $V$-band and several other bands for a set of stars; the ratio of the colour excesses is related to the slope of the dust extinction curve.  

Recently, spectral energy distribution (SED) fitting has become common, leading to modeling of the dust extinction curve using broadband photometric measurements.  However, this has mostly been limited to high-redshift galaxies, where one can observe rest-frame UV light using ground-based optical telescopes \citep[e.g.,][]{kriek13,price14,utomo14,zeimann15}.  In the case of nearby galaxies, there have been several different approaches.  \citet{conroy10b} use UV data from the Galaxy Evolution Explorer \citep[GALEX;][]{martin05} to measure changes in UV and optical colours as a function of galaxy inclination, and conclude that a 2175~\AA\ dust feature is necessary to explain their measurements.  \citet{hoversten11}, \citet{dong14}, and \citet{hutton15} use UV observations from the Ultraviolet/Optical Telescope \citep[UVOT;][]{roming00,roming04,roming05} on the Swift satellite \citep{gehrels04} to constrain the dust extinction curve in M81, the nucleus of M31, and M82, respectively.  \citet{demarchi16} use low resolution UV spectra from the Hubble Space Telescope to measure the dust curve in 30~Doradus in the Large Magellanic Cloud (LMC).  For each of these three latter studies, the galaxies' proximities mean that they could also measure spatial variability of the dust curve, a feat impossible at high redshift.

With this work, we are dramatically expanding our understanding of broad-scale dust properties in the SMC with  a method not feasible until now. There have been only a handful measurements of the shape of the dust extinction curve in the SMC, which are summarized in Table~\ref{prev_smc}.   These represent a total of 45 measurements, which includes many duplicates of the most useful stars.
These probe only a fraction of the SMC, yet the canonical ``SMC dust curve" is based on these measurements.  This paper directly addresses the clear need for more dust curve measurements in nearby galaxies.

In this paper, we use SED fitting to measure spatial variation of the dust extinction law in the Small Magellanic Cloud (SMC), utilizing UV observations from UVOT and archival optical and near-IR (NIR) imaging.  UVOT is uniquely suited to measure the dust extinction curve because of its three near-UV filters - $uvw2$ at 1928~\AA, $uvm2$ at 2246~\AA, and $uvw1$ at 2600~\AA\ - which are overlaid on the dust extinction curves plotted in Fig.~\ref{fig-dust}.  In particular, the $uvm2$ filter overlaps the 2175~\AA\ bump, so when the $uvm2$ flux is suppressed relative to those of $uvw2$ and $uvw1$, the degree of suppression traces the strength of the 2175~\AA\ bump.  Likewise, the amount $uvw2$ is extinguished compared to $uvw1$ helps to trace $R_V$, especially when combined with optical and NIR observations.  These capabilities mean that one can measure the spatial variability of the dust extinction curve on large scales with only broadband observations.  We note that an upcoming paper (Siegel et~al.\ 2016, in preparation) will use the Swift UV observations to derive the shape of the dust extinction curve for individual stars, whereas our approach models broader regions within the SMC.

\begin{table*}
\centering
\caption{Previous measurements of the dust curve in the SMC.  For the two references utilizing the colour excess method, the colour excesses of the stars are combined to make one measurement of the dust curve.}
\label{prev_smc}

\begin{tabular}{lcccc}
\hline
Reference & Method & Wavelength Coverage & Number of Stars & 2175~\AA\ Bump? \\
\hline
\citet{rocca81} & Pair & UV & 4 & Some \\
\citet{hutchings82} & Stellar Models & UV, Optical & 20 & N \\
\citet{lequeux82} & Pair & UV & 1 & Y \\
\citet{nandy82} & Pair & UV & 3 & Some \\
\citet{bromage83} & Compilation & UV, Optical & -- & N \\
\citet{prevot84} & Pair & UV, Optical & 7 & N \\
\citet{nandy84} & Colour Excess & Optical, NIR & 22 & -- \\
\citet{bouchet85} & Colour Excess & Optical, NIR & 23 & -- \\
\citet{thompson88} & Pair & UV & 5 & N \\
\citet{pei92} & Compilation & UV, Optical, NIR & -- & N \\
\citet{rodrigues97} & Pair & UV & 5 & Some \\
\citet{gordon98} & Pair & UV, Optical, NIR & 4 & Some \\
\citet{gordon03} & Pair & UV, Optical, NIR & 5 & Some \\
\citet{maiz12} & Stellar Models & UV, Optical, NIR & 4 & Some \\
\hline
\end{tabular}




\end{table*}


As a result of our modeling, we can also address the {recent ($<500$~Myr)} star formation history (SFH) of the SMC.  Measuring the SFH can help us understand the past interactions of the SMC, LMC, and Milky Way, and shed light on the evolution of dwarf galaxies in the local universe.
Most SFH studies of the SMC on large physical scales have used optical and IR light, primarily because of the lack of sufficiently deep wide-field UV observations of the SMC.  Previous work has found peaks in the SFR at
about 50~Myr \citep{harris04,indu11,rubele15}, 
300-600~Myr \citep{harris04,chiosi07,noel09,rezaeikh14}, 
1-3~Gyr \citep{harris04,chiosi07,noel09,piatti12,rubele15}, 
4-6~Gyr \citep{chiosi07,noel09,piatti12,cignoni12,weisz13,cignoni13,rezaeikh14,rubele15}, 
and
7-10~Gyr \citep{gardiner92,dolphin01,mccumber05,noel09,piatti12,weisz13}.
Many of the above papers argue that the peaks correspond to interactions between the SMC and LMC or Milky Way \citep[e.g.,][]{murai80,lin95},
or the accretion of low-metallicity gas \citep{yozin14}.

This paper is organized as follows.  In Section~\ref{sec-data}, we describe our UV, optical, and NIR data sets, and data reduction is discussed in Section~\ref{sec-data_red}.  We describe our SED modeling in Section~\ref{sec-modeling}.  We present our results in Section~\ref{sec-results} and discuss their significance in Section~\ref{sec-disc}.  We conclude in Section~\ref{sec-summary}.

%
%
%
%
%
%
%
%
%
%

%
%
%
%


\section{Data} \label{sec-data}

We use UV, optical, and near-IR imaging as the basis of our modeling.  The filters we use, in relation to the \citet{cardelli89} Milky Way dust extinction curve, are shown in the right panel of Fig.~\ref{fig-dust}.
In this work, we are modeling broad regions of star formation, so it is not necessary to maintain the high angular resolution of the multi-wavelength images.  In fact, for identifying large-scale overdensities, it is best that individual point sources are not prominent.  To this end, we use SWarp \citep[version 2.19.1;][]{bertin02} to simultaneously align each image and set the pixel scale to $10''$ (2.9~pc) for all images. \boldchange{SWarp does this by resampling the image at a scale smaller than the original pixels, rotated and offset as needed, then recombining the pixels to the desired scale and positioning. This has the effect of both re-centering and rebinning the image.} 
The images are centred at $\alpha = 0^h 55^m 19.77^s$, $\delta = -72^\circ 47^m 0.92^s$ and have dimensions of $2.45^\circ \times 2.01^\circ$ (2.6~kpc $\times$ 2.1~kpc).  Most of the imaging described below does not cover this entire area, but except for a few small regions, the limiting footprint is our UVOT mosaic.

\subsection{Ultraviolet}

The SUMaC  (Swift Ultraviolet survey of the Magellanic Clouds)  program is the first comprehensive \boldchange{multi-filter NUV survey covering the inner} Magellanic Clouds\footnote{\boldchange{GALEX has observed the entirety of the Magellanic Clouds in the NUV filter, and analysis is ongoing (\citealt{simons14}; Seibert \& Schiminovich, in preparation)}}.
Initiated as a team project, it provides three-filter NUV coverage of the cores
of the Small and Large Magellanic Clouds, along with an X-ray survey utilizing the X-Ray Telescope \citep{burrows05}, 
to match previous surveys performed in the optical and IR.
Specific scientific goals of the program \boldchange{were to}:

\begin{enumerate}

\item[(i)] Investigate the NUV properties of star forming regions in the Clouds and the relationship between
star formation rate indicators in the low metallicity environments of the Clouds,

\item[(ii)] Identify and study hot stars, and blue hook stars and Wolf-Rayet stars in particular, in the
low metallicity environment of the Clouds,

\item[(iii)] Constrain the contribution of blue hook stars to the reionization of the universe,

\item[(iv)] Trace the recent ($<$500~Gyr) star formation history of the Clouds to
greater precision that can be done with optical surveys,

\item[(v)] \boldchange{Compare} the star formation history to recent dynamical interactions between
the two Clouds and between the Clouds and the Milky Way,

\item[(vi)] Improve UV stellar evolution isochrones, for both stellar models and spectral synthesis models in the UV,

\item[(vii)] \boldchange{Measure} the NUV extinction curve across the face of the Clouds and search for insights
into the physical cause of the 2175~\AA\ bump,

\item[(viii)] \boldchange{Identify} background QSOs as reference points for future studies of Cloud extinction
and absolute proper motions,

\end{enumerate}

\noindent
In this paper, we focus on \boldchange{topics} (iv), (v), and (vii); the remaining goals will be addressed in future work.

The SMC was observed in a staggered pattern of 50 tiles, each $17 \times 17$ arcminutes, with a few
previous observations used to patch the coverage.  All observations were taken in $2 \times 2$ binned mode with a pixel scale of $1.0''$.  Observations began on 2010 September 26 and ended on 2013 November 6, with the majority of the observations made between May 2011 and December 2011.  Typical exposure times were 1~ks per filter in the $uvw2$, $uvm2$ and $uvw1$ filters.  In Table~\ref{filter_data}, we list the properties of the three filters, the median exposure times, areas of the images, \boldchange{and the 3$\sigma$ limiting surface brightnesses}.  
\boldchange{For point sources, the 50\% detection limit is typically around 18.7~AB~mag, but this varies considerably with background and crowding (Siegel et al., in preparation).}
For a detailed discussion of the filters, as well as plots of the responses, see \citet{poole08} and updates in \citet{breeveld11}.

The LMC was observed in a staggered pattern of 171 tiles.  Observations
began on 6 July 2011 and ended on 2 April 2013, with the majority of the observations
made between May 2011 and December 2011 and between October 2012 and April 2013. Typical exposure times
were also 1~ks per filter in the $uvw2$, $uvm2$ and $uvw1$ filters.  Analysis of the LMC will be presented in a future paper.

\begin{figure*}
	\centering
	\includegraphics[trim = 0mm 0mm 0mm 0mm, clip=true, height=0.33\textwidth]{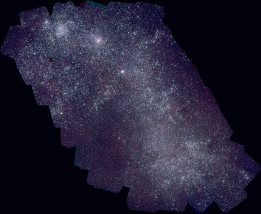}
	~
	\includegraphics[trim = 0mm 0mm 0mm 0mm, clip=true, height=0.33\textwidth]{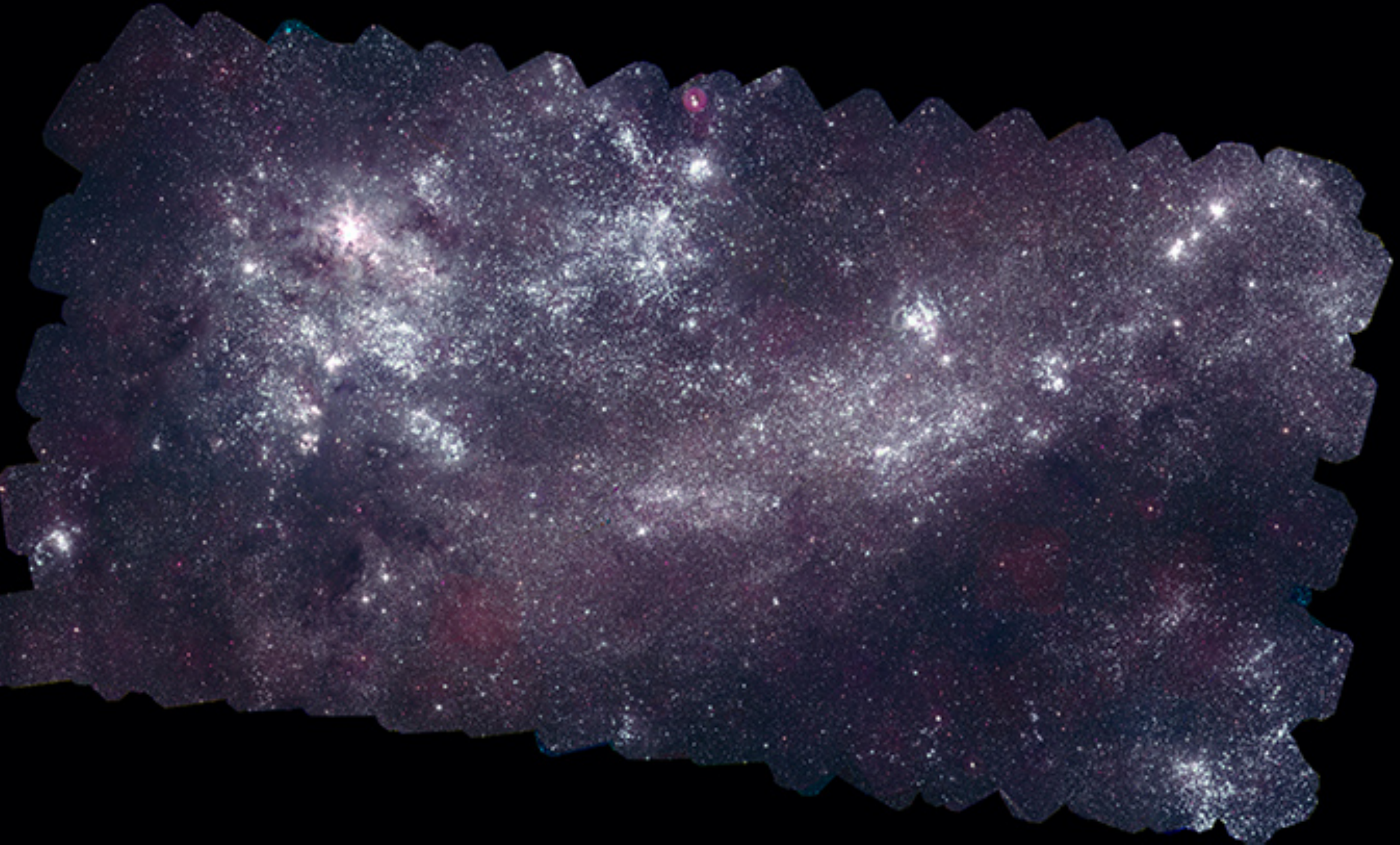}
	\caption{False colour UVOT images of the SMC (left) and LMC (right) with $uvw2$ (blue), $uvm2$ (green), and $uvw1$ (red).  The SMC image is about 2.3$^\circ$ (2.4~kpc) across, and the LMC image is 4.4$^\circ$ (3.9~kpc) across.  North is to the top and east is to the left.}
	\label{fig-pretty}
\end{figure*}

For both galaxies, the automated aspect solution failed on numerous occasions due to the exceptionally crowded field.  This required extensive manual correction of the images to a consistent astrometric system \citep[see details in][]{siegel14}.  Fully mosaicked colour images were released to the public in June of 2013, and are shown in Fig.~\ref{fig-pretty}.  Mosaicked and individual FITS files are available upon request.

%

\begin{table*}
\centering
\caption{\swift/UVOT Observations of the SMC. The filters' central wavelengths (assuming a flat spectrum), FWHMs, and image PSFs are from \citet{breeveld10}. }
\label{filter_data}


\begin{tabular}{lccccccc}
\hline
Filter & Central Wavelength & FWHM & PSF FWHM & Median & Area & \boldchange{3$\sigma$ Limiting $\mu$} \\ 
  & (\AA) & (\AA) &   & Exposure (s)  & (deg$^2$) & \boldchange{(AB mag/arcsec$^2$)} \\
\hline
$uvw2$ & 1928 & 657 & $2.92''$ & 1202 & 2.759 & 22.87 \\
$uvm2$ & 2246 & 498 & $2.45''$ & 1127 & 2.758 & 22.27 \\
$uvw1$ & 2600 & 693 & $2.37''$ & 1077 & 2.752 & 22.54 \\
\hline
\end{tabular}

\end{table*}



\subsection{Optical}

We use optical imaging from \citet{massey02}.  The data were taken at the Curtis Schmidt telescope at CTIO in the Harris \textit{UBVR} filters \citep{massey00}. The imaging covers six fields, each $1.3^\circ \times 1.3^\circ$, of which four overlap with our UVOT imaging.
The images have a resolution of between $0.9''$ and $1.3''$ (0.61~to 0.87~pc at the distance of the SMC); the PSF is undersampled, but we do not use the full-resolution data in our analysis.
The $U$-band calibration is dependent upon the stars' surface gravities (see the discussion in \citealt{massey02} for details), so we discard it from analysis.
For the remaining \textit{BVR} filters, their approximate $5\sigma$ depths for point sources are 17.0/17.2/16.5 AB~mag, respectively, though it should be noted that these vary by up to 0.4~mag depending on position \boldchange{and stellar crowding}.





\subsection{Infrared}

The SMC was observed as part of the Two Micron All-Sky Survey \citep[2MASS;][]{cohen03}.  We acquired mosaics using the online interface of the Montage software package.  The sky background on any given 2MASS image varies considerably over the course of the exposure; although Montage attempts to correct for this variation, there are still discontinuities where the exposures overlap in the mosaics.  Only the J-band image have smooth enough variation to be easily correctable, so we discard the H and K$_\text{s}$ images.  The 2MASS imaging tiles in the SMC have an exposure time of 7.8~s with small overlapping areas, \boldchange{and the 5$\sigma$ point source depth is 18.9~AB~mag}.



The SAGE (Surveying the Agents of Galaxy Evolution) SMC program \citep{gordon11} includes imaging of the SMC and Magellanic Bridge with \textit{Spitzer} in the IRAC 3.6, 4.5, 5.8, and 8.0~$\mu$m bands and the MIPS 24, 70, and 160~$\mu$m bands.  
We only utilize the 3.6~$\mu$m observations for two reasons: (1) beyond 3.6~$\mu$m, the light is dominated by dust emission, so longer wavelengths do not add any constraints to the shape of the dust extinction curve, and (2) our modeling (see \S\ref{sec-modeling}) does not account for emission from dust, though we will address dust emission in future work.
The 3.6~$\mu$m imaging has $1.7''$ resolution and has an exposure time of 185~s along the bar and wing and 41~s elsewhere.  The point source catalog has a 5$\sigma$ depth of  19.8~AB~mag over the whole survey area.  




\section{Data Reduction} \label{sec-data_red}


We map the properties of the SMC in two ways.  We use Source Extractor \citep{bertin96} to identify individual star-forming regions, from which we extract photometry and model the SEDs.  In addition, we bin the images into large $200''$ (58~pc) pixels, which we also model.
We adopt a distance modulus of 18.91 (60~kpc) for the SMC \citep{hilditch05}, though we note that our final results do not depend on the distance.

\subsection{Background}

Before photometering anything in the SMC, it is necessary to calculate the contribution of diffuse background light.  We use different procedures for the star-forming regions and large pixels. 

For the star-forming regions, our goal is to isolate the clumpy material associated with recent star formation.  Therefore, we consider the background to be any light that is diffuse.  To remove this diffuse light,  we use a circular median filtering technique following \citet{hoversten11}.
We calculate the background \boldchange{in each bandpass} using the images with $10''$ (2.9-pc) pixels.  For a given pixel, we calculate the median value of pixels within a range of radii.  \citet{pleuss00} used HST imaging of HII regions in M101 to show that typical HII regions range between 20~pc and 220~pc in diameter; we therefore measure the median within circles of radius 10, 15, 20, 25, 35, 50, 65, 80, 95, and 110~pc about each pixel.  We then take the minimum of these medians as the pixel's background, without exceeding the pixel's original value.
This procedure has several advantages.  First, because the sizes and structures of star-forming regions vary, it ensures that a reasonable background is calculated for each one.  Second, the background map preserves the detailed shapes of heavily extinguished areas, excluding the possibility of over-subtraction.

For the large pixels, we want to model all light within the pixel, including light from older stellar populations.  This facilitates a more direct comparison to previous work that only includes optical/IR light.  The background is then composed of any large scale instrumental or sky background.
To remove this background, before binning, we subtract the mode of each image.  Using the mode, rather than the mean or median, ensures that the background estimate is not affected by emission from the SMC.  Any pixels that are less than the mode are set to zero.  In areas with especially low count rates in a given filter (especially the southeast part of the SMC), some pixels had near-zero count rates, making their photometry unreliable; these data points were removed prior to modeling.

\boldchange{For the UVOT imaging, the survey area is almost entirely composed of emission from the SMC, so the image mode is not representative of the true background value.  To calculate the background, we utilized an archival UVOT pointing $2.6^\circ$ offset from our survey, centered at the coordinates $0^\text{h} 27^\text{m} 25.8^\text{s}$, $-71^\circ 22' 30.7''$, with a total exposure time of 9600s in $uvw2$, 10000s in $uvm2$, and 5700s in $uvw1$.  We rebinned the images to the same $10''$ pixel scale and followed the same procedure as above to calculate the mode background value.}

\subsection{Star-forming Regions} \label{sec-data_red-sfr}

Regions of recent star formation in the SMC have been identified in many different ways:
H$\alpha$ line emission \citep[e.g.,][]{kennicutt86,lecoarer93,kennicutt08},
dust emission \citep[e.g.,][]{lawton10,gordon14},
and
radio observations of hydrogen and molecular clouds \citep[e.g.,][]{stanimirovic99,bot10}
are the most common.  Here we take a different approach by using UV light, which is directly emitted by massive stars, complementing the observations of reprocessed UV photons at other wavelengths.

\begin{figure*}
	\centering
	\includegraphics[trim = 35mm 58mm 25mm 75mm, clip=true, width=0.75\textwidth]{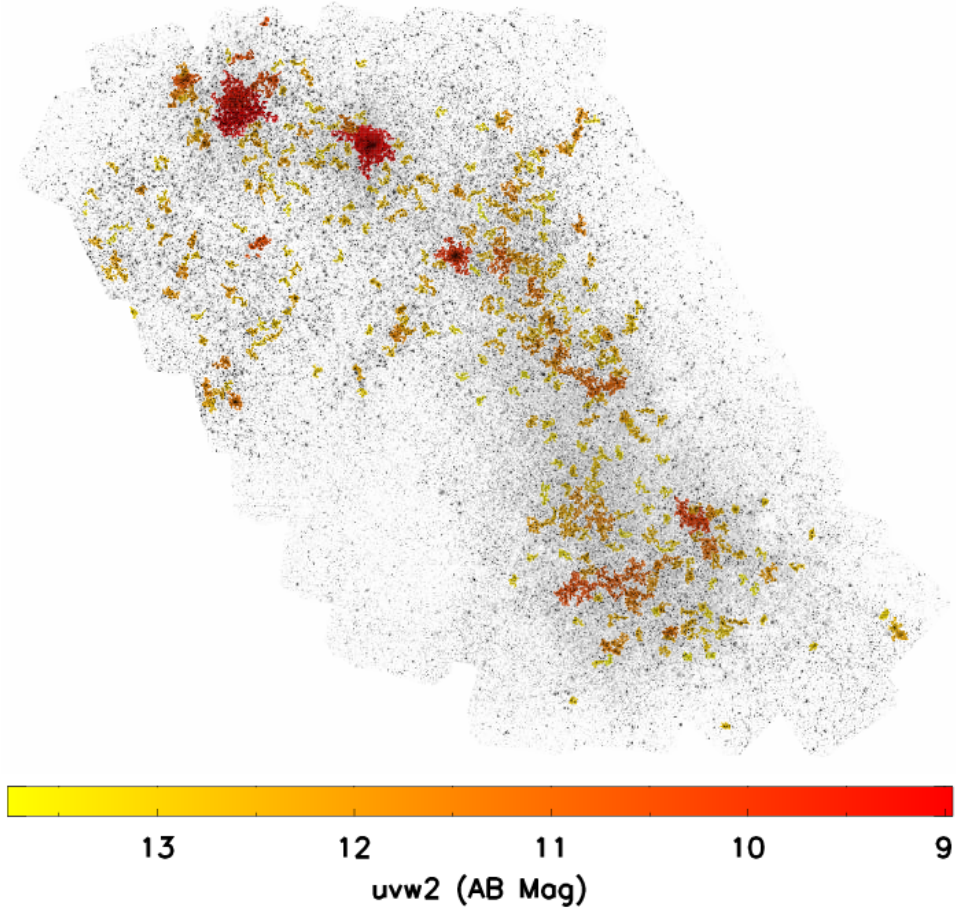}
	\caption{Map of the 338 star-forming regions overlaid on the $uvm2$ image, color coded by $uvw2$ brightness.}
	\label{fig-sf_map}
\end{figure*}

We use Source Extractor \citep[SE; version 2.5.0;][]{bertin96} to identify star-forming regions in the UVOT $uvw2$ background-subtracted image.  We choose this filter because it is the bluest available, thus tracing the most massive young stars.  The $uvw2$ filter does have a red leak, but the transmission isn't significant beyond $\sim$3000\AA\ \citep{siegel14,breeveld10,brown10}.  We require that regions be composed of at least 30 pixels \boldchange{(each pixel rebinned to $10''$ as described in Section~\ref{sec-data})}, which is an area of 250~pc$^2$.  SE has a known problem in which it can identify pixels in non-contiguous regions as belonging to the same region; to correct for this, we follow the method described in Appendix~A of \citet{hoversten11}.
After this correction, we identify 338 star-forming regions, \boldchange{shown in Fig.~\ref{fig-sf_map}}.

Using the SE-defined regions, we extract photometry from each of the background-subtracted images.  While SE can do this, it cannot properly propagate errors, so we do the photometry manually.  This entails retrieving the pixels corresponding to the SE regions from the original image, the background image, and the exposure map for each bandpass.  The photometric uncertainties take into account the Poisson errors from both the original image and the background.  We set a minimum uncertainty of 0.05~mag for each star-forming region. 
\boldchange{The signal-to-noise of the regions in the $uvw2$ detection band ranges from 260 to 3600 with a median of 540.}

This method of detecting star-forming regions introduces selection biases for age and dust properties.  We address these in detail using model SEDs in Section~\ref{sec-sf_detect}.
In order to make broad statements about the galaxy, we also break up the SMC images into large pieces for modeling, described below.


\subsection{Pixel-by-Pixel} \label{sec-pix}

Modeling the entire map of the SMC somewhat alleviates the selection biases inherent in detecting star-forming regions.  However, due to computational constraints when modeling the SEDs, the resolution is necessarily much coarser.  In addition, the broad combination of distinct epochs of star formation smooths over the detailed star formation history, which affects the final results.

To make the map, we re-bin the background-subtracted images into $200''$ (58~pc) pixels.  Since the images used to extract star-forming regions are $10'' \times 10''$ pixels, this is simply doing $20 \times 20$ binning.  
In some cases, the large pixels are partially comprised of small pixels beyond the edge of the imaged region.  If over 10\% of these small pixels are unusable from any bandpass (i.e., CCD imperfections), the large pixel is discarded from analysis.
The UV and optical images cover the smallest fields of view, so the outer boundary is effectively determined by these images.  This procedure results in 775 large pixels across the SMC.

We extract photometry in each bandpass in much the same way as for the star-forming regions.  For each large pixel, we would optimally sum the fluxes of the constituent pixels, but up to 10\% of those pixels could be masked.  Therefore, we take the mean of the non-masked pixels and multiply by the total large pixel area (400 small pixels).  As before, we set a minimum uncertainty of 0.05~mag.

\subsection{Foreground Dust}

It is necessary to account for absorption by dust from the Milky Way along our line of sight.  In order to correct our flux measurements, we use the \citet{schlegel98} dust maps.  However, for nearby galaxies (including the SMC), \citet{schlegel98} advise that the dust measurements are unreliable due to contamination from the galaxies themselves.  In the vicinity of the SMC, the amount of dust from the Milky Way is quite low and appears to not vary on small angular scales.  Therefore, we use the \citet{schlegel98} measurement of the median dust in an annulus around the SMC.  The resulting dust extinction is $E(B-V) = 0.037$, which corresponds to $A_V = 0.11$ for $R_V = 3.1$.  We assume the \citet{cardelli89} Milky Way dust extinction curve - which has a 2175~\AA\ bump - to correct each of the fluxes.  The dust-corrected photometry is listed in Table~\ref{tab-sf_phot} for the star-forming regions and Table~\ref{tab-pix_phot} for the large pixels.

\input{table3.tex}

\input{table4.tex}


\section{Modeling} \label{sec-modeling}

We model the spectral energy distributions of each star-forming region and \boldchange{large} pixel by comparing our data to a grid of models.  We create these models using the PEGASE.2 spectral synthesis code \citet{fioc97}.  
We use a \citet{salpeter55} initial mass function spanning 0.1~\msun\ to 120~\msun.  We utilize the stellar evolution tracks assembled by \citet{fioc97}, which are a combination of many observed and theoretical spectra.  Our data do not strongly constrain the metallicity, so we assume a metallicity of $0.25Z_\odot$, which corresponds to ages of less than 1~Gyr in the SMC age-metallicity relation \citep[e.g.,][]{harris04}.  Nebular emission lines are included.

The grid includes spectra for ages of 1~Myr to 13~Gyr.
We assume that for the isolated individual star-forming regions, a single instantaneous starburst is a reasonable star formation history.  This also reduces the total number of physical parameters to fit, which enables better constraints on the remaining parameters.  For the pixels, on the other hand, each one is likely accounting for a variety of star formation histories; for these, we also fit an exponentially decreasing star formation history (SFH) with time scales ($\tau$) from 110~Myr to 3.5~Gyr.

Once the PEGASE model spectra are generated, we apply a grid of extinction laws.  These are parametrized following \citet{cardelli89}, varying the dust extinction curve slopes ($R_V$) from 1.5 to 5.5 and 2175~\AA\ bump strengths from 0 to 2 (where 1 is the strength found in the Milky Way).  For each combination of $R_V$ and bump strength, we scale by a dust attenuation ($A_V$) of 0 to 7 magnitudes.  Finally, from each spectrum in this multi-dimensional grid, we extract the model fluxes for each filter using the published filter transmission curves.

We find the best-fitting physical parameters using \verb=emcee= \citep{foreman-mackey13}, a Markov-chain Monte Carlo (MCMC) sampling code.  A significant advantage of the MCMC technique is that it can reveal degeneracies between physical parameters.  This is very important because many statistical techniques require the assumption of uncorrelated uncertainties.  Also, many other fitting methods assume Gaussian uncertainties, but never test if that assumption is correct.  With MCMC, it is trivial to test for both uncertainty symmetry and Gaussianity.  In addition, the MCMC method searches a wide parameter space, so it can discover and quantify multi-modal distributions of parameter values.

Using the \verb=emcee= code, we fit for the age, SFH time scale $\tau$ (for the pixels only), dust parameters ($A_V$, $R_V$, and bump strength), and processed mass, which includes stars and remnants.  The mass is simply a normalization, but by fitting for it, we can measure its uncertainty and any degeneracies with other parameters.  PEGASE also has prescription for dividing the processed mass into its constituent stellar mass and mass of stellar remnants.  This prescription depends on age and $\tau$, so we derive the masses of these components after fitting for the other physical parameters.  It is also important to note that this is a closed box model, so we cannot account for if the SMC accretes or is stripped of material \citep[e.g., the Magellanic Stream,][]{gardiner96}.

For each step in the \verb=emcee= code, we calculate the model flux by interpolating the model flux grid, and then compare our photometry (from one star-forming region or one pixel) to the model.  The log likelihood for this comparison is
\begin{equation}
\ln L = -\frac{1}{2} \chi^2
=
- \frac{1}{2}
\sum_i \frac{ (F_{\text{obs,} i} - F_{\text{model,} i} ) ^2 }
{ (\delta F_{\text{obs,} i} )^2 } ,
\end{equation}
where $F_\text{obs}$ is the observed flux, $F_\text{model}$ is the modeled flux, and $\delta F_\text{obs}$ is the uncertainty in the observed flux.

We run the MCMC process with 2000 chains for star-forming regions and 4000 chains for large pixels, starting at random locations in $n$-dimensional parameter space.  Because the parameter space is so large, this large number of chains ensures that all parts of parameter space are fully explored.
We knew from preliminary runs that the SMC has only a small amount of dust, so to reduce the number of steps before convergence, we limit the starting $A_V$ to between 0 and 1.0~mag.   
(The chains can still explore the full parameter space.)  

\begin{figure}
	\centering
	\includegraphics[trim = 0mm 5mm 0mm 10mm, clip=true, width=0.95\columnwidth]{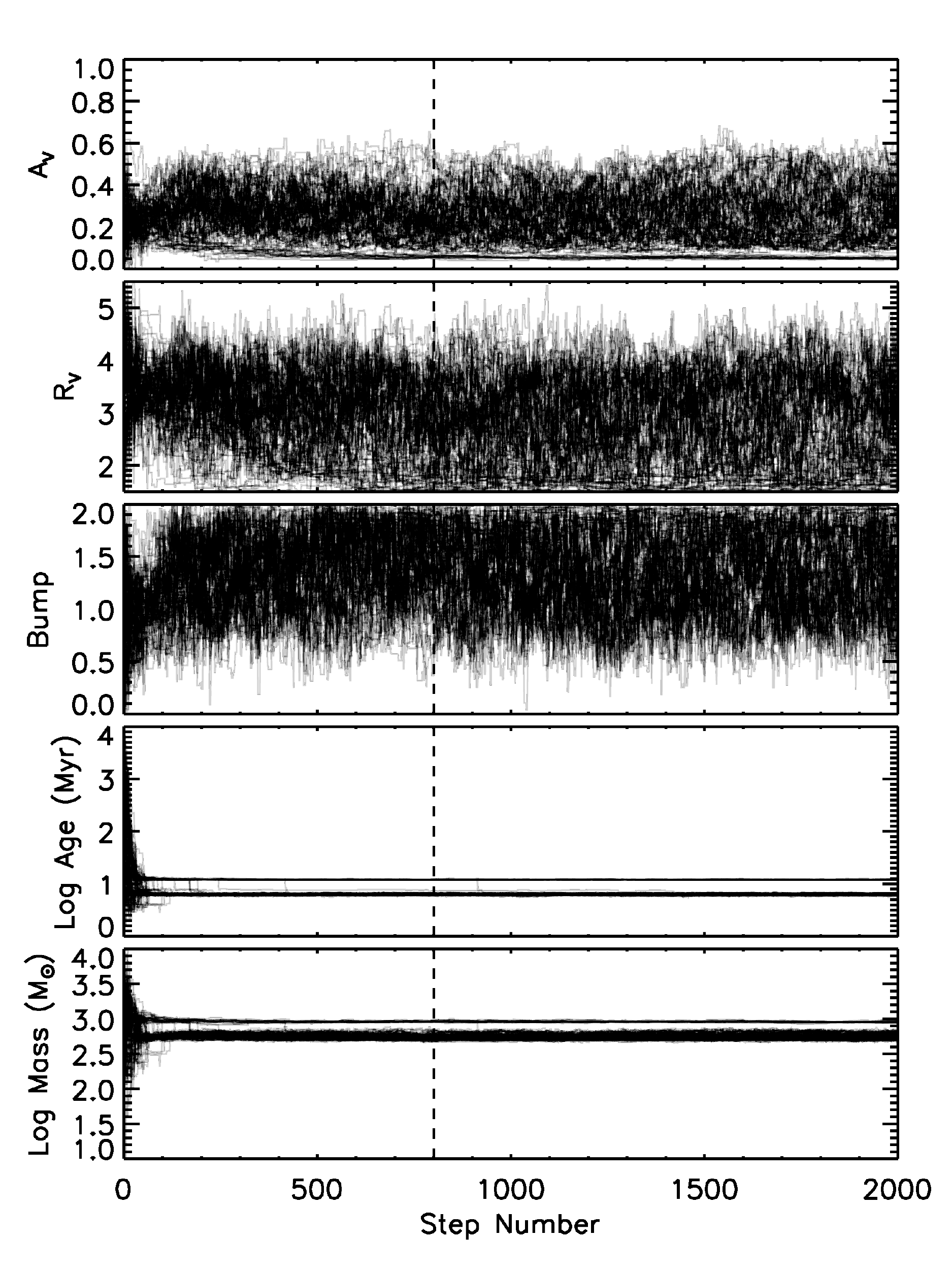}
	\caption{200 randomly-selected chains (out of 2000 total chains) plotted for each physical parameter for star-forming region 165.  The chains appear to stabilize by about step 400, but we set a more conservative burn-in of 800 steps, marked by the vertical dashed line.}
	\label{fig-chains}
\end{figure}

Each chain is run for 2000 steps.  Convergence typically occurs by 500 steps, but we set a more conservative burn-in of 800 steps, meaning that the final parameter values are derived from only the last 1200 steps of each chain.  A set of chains for one star-forming region is shown in Fig.~\ref{fig-chains}.
We combine all 2.4~or 4.8~million points (1200 steps from 2000 or 4000 chains) and remove any severe outliers.  Each point is made up of five (for star-forming regions) or six (for pixels) parameter values.  

\begin{figure*}
	\centering
	\includegraphics[trim = 30mm 65mm 30mm 100mm, clip=true, width=0.45\textwidth, page=1]{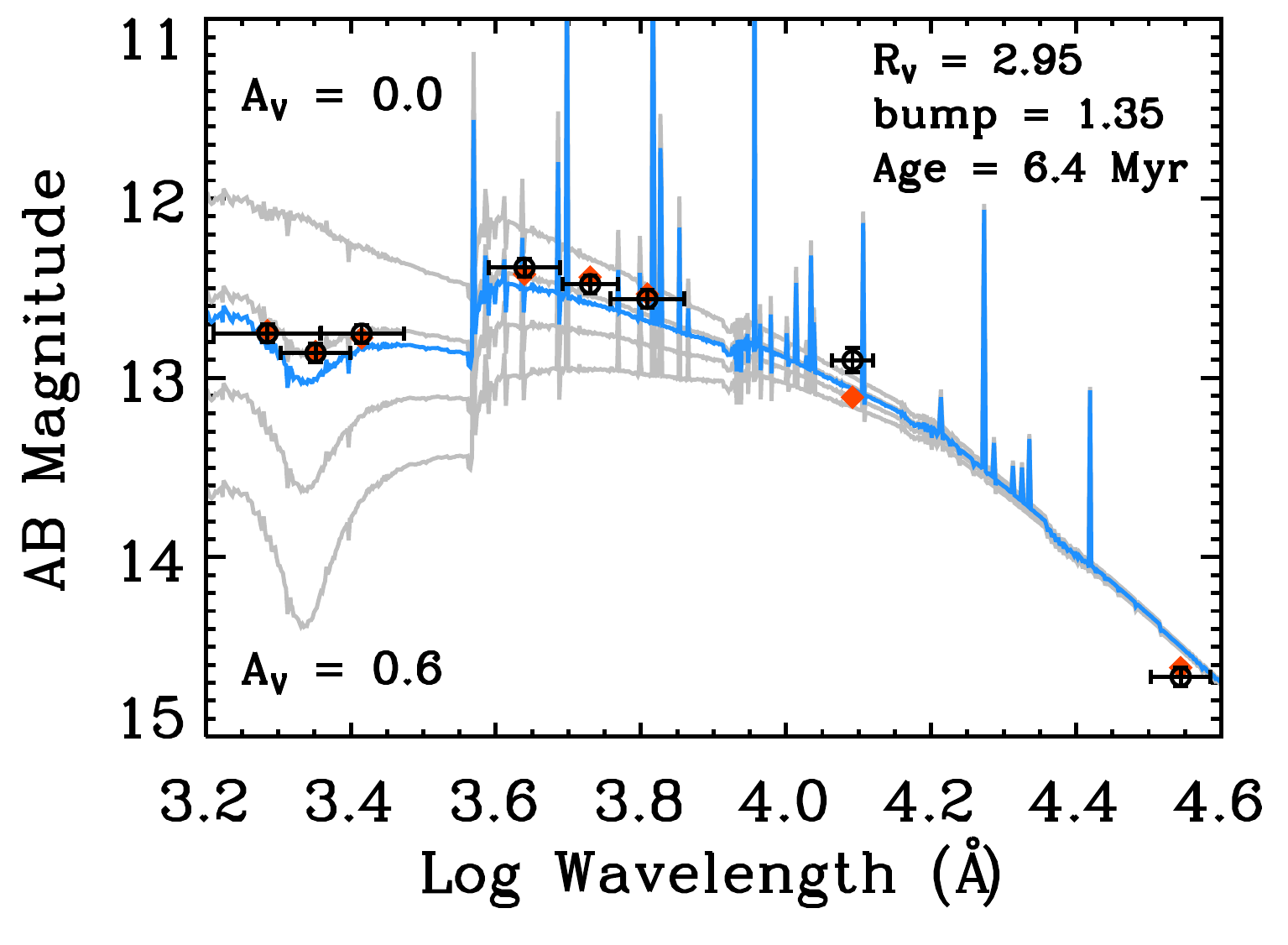}
	\includegraphics[trim = 30mm 65mm 30mm 100mm, clip=true, width=0.45\textwidth, page=2]{fig5.pdf}
	\includegraphics[trim = 30mm 65mm 30mm 100mm, clip=true, width=0.45\textwidth, page=3]{fig5.pdf}
	\includegraphics[trim = 30mm 65mm 30mm 100mm, clip=true, width=0.45\textwidth, page=4]{fig5.pdf}
	\caption{Demonstration of how changing the physical parameters effects the SED.  The photometry (black circles) is for star-forming region 165, with the best-fitting spectrum (blue) and corresponding best-fitting magnitudes (red diamonds), which corresponds to $A_V = 0.24$~mag, $R_V = 2.95$, a bump strength of 1.35, and an age of 6.4~Myr.  \boldchange{The signal-to-noise of region 165 is near the median for star-forming regions.  For each parameter that is varied, the values of the fixed parameters are noted at the top right of the plot.}
	\textit{Top left}: Varying $A_V$ from 0 to 0.6 magnitudes.  Changes in $A_V$ have the largest effect at shorter wavelengths, with very little change in the infrared.
	\textit{Top right}: Varying $R_V$ from 1.5 to 5.5.  The largest variations are in the UV, with much smaller changes at optical wavelengths.
	\textit{Bottom left}: Varying the bump strength from 0 to 2.  The primary effect is around the $uvm2$ filter.
	\textit{Bottom right}: Varying the age from 4 to 8~Myr.  At these young ages, the spectrum is changing dramatically at all wavelengths.}
	\label{fig-spec}
\end{figure*}

The degree to which each photometric point constrains the models is shown in Fig.~\ref{fig-spec}.
The UV data plays a vital role in constraining $R_V$ and the bump strength.  In particular, it's worth noting that while differences in $R_V$ slightly affect the optical brightness, the effect in the UV is considerably larger, so UV data is important to put tight limits on the $R_V$ values.
The combined optical and UV together put constraints on $A_V$; without optical data, it would be difficult to tell whether variation in the UV data was due to differences in $A_V$ or $R_V$.
Finally, varying the age of the stellar population has an effect on both the shape and the normalization of the entire SED, so measurements from UV to near-IR contribute to our modeled ages.

\section{Detection of Star-Forming Regions} \label{sec-sf_detect}

There are several important considerations for identifying star-forming regions.  First, as a region of star formation ages, it gets fainter, so its emission of UV light decreases over time.  The older star-forming regions in the SMC will then be less likely to be detected using SE.
Second, the dust properties of a region will determine how easily the region is detected.  Areas of star formation with large amounts of obscuring dust will be less readily detected in the UV.
Third, star-forming regions evaporate over time, and a more dispersed group of stars are less likely to be identified by SE.

\begin{figure*}
    \centering
    \includegraphics[trim = 30mm 55mm 15mm 60mm, clip=true, scale=0.32, page=1]{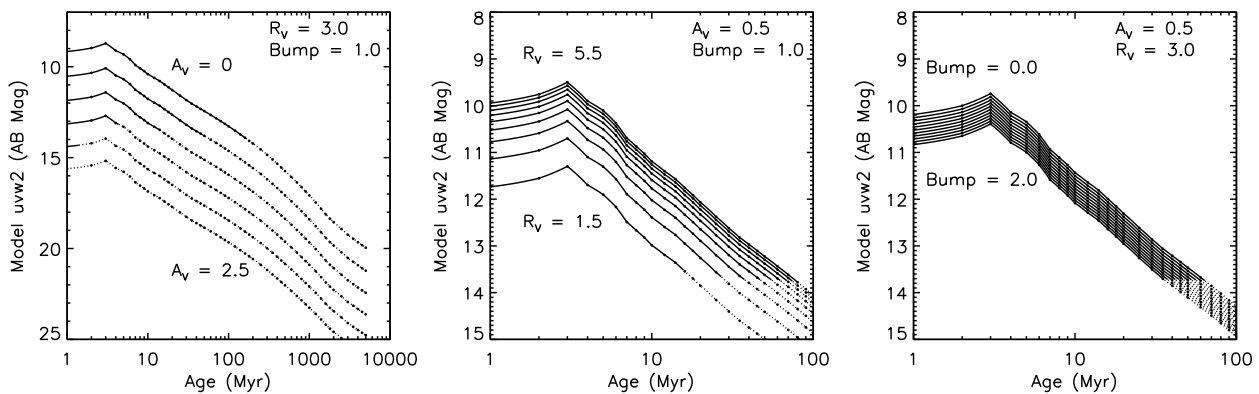}
    \includegraphics[trim = 30mm 55mm 15mm 60mm, clip=true, scale=0.32, page=2]{fig6.pdf}
    \includegraphics[trim = 30mm 55mm 15mm 60mm, clip=true, scale=0.32, page=3]{fig6.pdf}
    \caption{Time evolution of the $uvw2$ magnitude as a function of the dust parameters.  All curves are for a starburst with a mass of 5000~\msun\ at 60~kpc.  \boldchange{The line changes from solid to dotted at the age when a concentrated, low-background region (see text for definitions) becomes undetectable.}
    Note the changing vertical and horizontal scales in each panel. 
    \textit{Left}: Increasing $A_V$ from 0 to 2.5 with $R_V = 3.0$ and a bump strength of~1.  The dust attenuation has a pronounced effect on the visibility of the $uvw2$ light.
    \textit{Middle}: Decreasing $R_V$ from 5.5 to 1.5 (increasing the slope of the dust curve) with $A_V = 0.5$ and a bump strength of~1.  Changing $R_V$ has a small effect on $uvw2$ at high values, but it becomes more significant as $R_V$ becomes lower.
    \textit{Right}: Increasing the 2175~\AA\ bump strength from 0.0 to 2.0 with $A_V = 0.5$ and $R_V = 3$.  The bump strength has only a small impact on the modeled $uvw2$ magnitude.}
    \label{fig-w2_evolution}
\end{figure*}

We address the first two issues in Fig.~\ref{fig-w2_evolution}.  The model light curves are generated using the models described above.
In the figure, one can initially see that the $uvw2$ magnitude decreases with time; in flux units, $F_{uvw2} \propto \text{(age)}^\alpha$ with $\alpha \approx 3.5$.
We can quantify the effect of this for a 5000~\msun\ region (as used in Fig.~\ref{fig-w2_evolution}) with no dust and a signal-to-noise representative of that of the measured star-forming regions.  If we assume a low (high) background \boldchange{of 24.1 (22.9) mag/arcsec$^2$}, a concentrated region \boldchange{with an area of 30~pixels} will be detectable until an age of 160~Myr (120~Myr), while a diffuse region \boldchange{with an area of 120~pixels} will drop below the threshold by an age of 110~Myr (70~Myr).  These ages scale linearly with the total mass.  

\boldchange{One can also consider the limiting magnitude at which a given star-forming region will be detected.  A 5000~\msun\ region with $A_V = 0.5$ and $R_V = 3.0$ will have a wide range of minimum brightnesses depending on the size and background levels.  Using the same definitions of region size and background as above, a concentrated region with a low (high) background has a limiting AB magnitude of 13.70 (13.35), and a diffuse region has a limiting AB magnitude of 13.20 (12.55). }


In the first panel of Fig.~\ref{fig-w2_evolution}, the total dust ($A_V$) has a significant effect on the measured brightness of a star-forming region.  For $R_V = 3$, each magnitude increase in $A_V$ corresponds to a decrease of approximately 2.7 magnitudes in $uvw2$.  As such, obscuring a star-forming region will make it even more difficult for SE to identify.  When $R_V$ is small, this effect is even stronger; the middle panel of Fig.~\ref{fig-w2_evolution} demonstrates that for a given $A_V$, a steepening of the extinction curve has an increasingly large effect on the measured $uvw2$ magnitude.  Changing the strength of the dust bump, as seen in the right panel, contributes very little to the variation in $uvw2$ obscuration.

Evaporation of detected star clusters is also a possible concern.  
Following \citet{lada03}, for clusters with masses of $\sim$200~\msun\ (2000~\msun), the evaporation time scale is $\sim$$1.5 \times 10^8$~yr ($9 \times 10^8$~yr).  As calculated in \S\ref{sec-results}, 99\% of modeled masses are above 200~\msun\ (51\% above 2000~\msun) and 97\% of modeled ages are younger than $1.5 \times 10^8$~yr (100\% younger than $9 \times 10^8$~yr).  This suggests that cluster dissipation is a negligible factor in the detection of star-forming regions.
\boldchange{More recently, there has been significant discussion and debate about whether cluster disruption is mass-dependent \citep{lamers05} or mass-independent \citep{whitmore07}, including whether there is a dependence on cluster environment \citep[e.g.,][]{bastian11,chandar14}.  More investigation is needed before we can assess the detailed impact of these proposed scenarios on our results. }

Properly accounting for these selection criteria is very difficult (if not impossible).  Therefore, we caution that the physical parameters derived from the star-forming regions (Section~\ref{sec-results}) should not be used as a global representation of the SMC.  Many regions of parameter space are inaccessible, particularly combinations of lower mass, larger dust extinction, and older age.


\section{Results} \label{sec-results}

\begin{figure}
	\centering
	\includegraphics[trim = 5mm 35mm 0mm 35mm, clip=true, width=0.98\columnwidth]{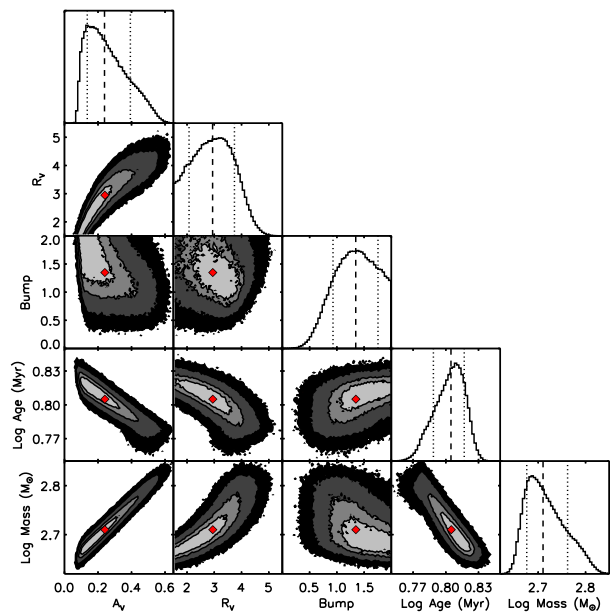}
	\caption{For star-forming region 165, each set of physical parameters in the post-burn-in chain plotted against each other.  Contours represent 0.5, 1, 2, and 3 sigma, \boldchange{and red diamonds mark the best fit values}.  It is easy to see which parameters have strong degeneracies with others, i.e, their uncertainties are correlated.  In particular, $A_V$, $R_V$, the age, and the stellar mass tend to be degenerate with each other.  Along the diagonal are histograms for each parameter, with the median (best-fitting) value marked with a dashed line and $\pm1\sigma$ marked with dotted lines.}
	\label{fig-triangle}
\end{figure}

In Fig.~\ref{fig-triangle}, we show an example of the parameter space explored by post-burn-in chains modeling a star-forming region.  The steep slopes in the contour plots show that there are significant degeneracies between  $A_V$, $R_V$, age, and stellar mass, meaning that their uncertainties are correlated.  Physically, this makes sense: the model UV fluxes can be decreased by increasing $A_V$, increasing $R_V$, increasing the age (or decreasing if below 3~Myr), or lowering the mass.
Similarly, the optical/IR fluxes can be decreased by increasing $A_V$ or lowering the mass.  \boldchange{We do note, however, that both the age and mass are very well constrained, even with these degeneracies.} The bump strength is not notably degenerate with any of the other quantities; this is especially important because it means the bump strength is well constrained by our data.

The best-fitting values and uncertainties for the star-forming region can also be seen along the diagonal in Fig.~\ref{fig-triangle}.  The histograms are a coarse probability distribution function for each parameter, so we can define the best fits as the 50th percentiles, with $1\sigma$ uncertainties from the 16th and 84th percentiles.  The histograms in Fig.~\ref{fig-triangle} are not symmetric or Gaussian, and neither are they for most of the other star-forming regions and large pixels.  \boldchange{We also find that the probability distributions tend to be unimodal.}  We compile the best-fitting physical parameters and their uncertainties in Table~\ref{tab-sf_prop} (star-forming regions) and Table~\ref{tab-pix_prop} (large pixels).


\input{table5.tex}

\begin{figure*}
	\centering
	\includegraphics[trim = 30mm 30mm 25mm 75mm, clip=true, width=0.3\textwidth]{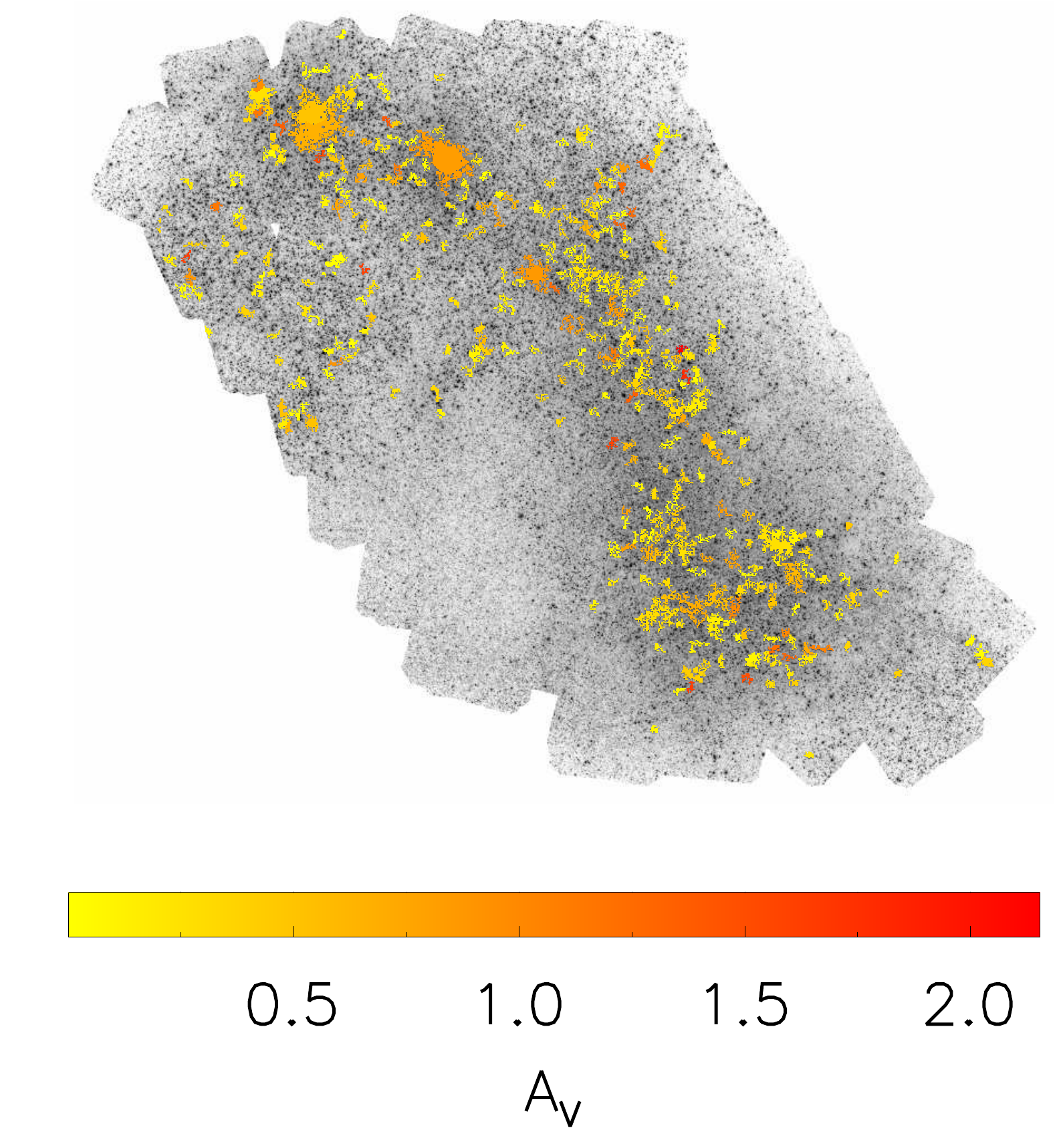}
	\includegraphics[trim = 30mm 30mm 25mm 75mm, clip=true, width=0.3\textwidth]{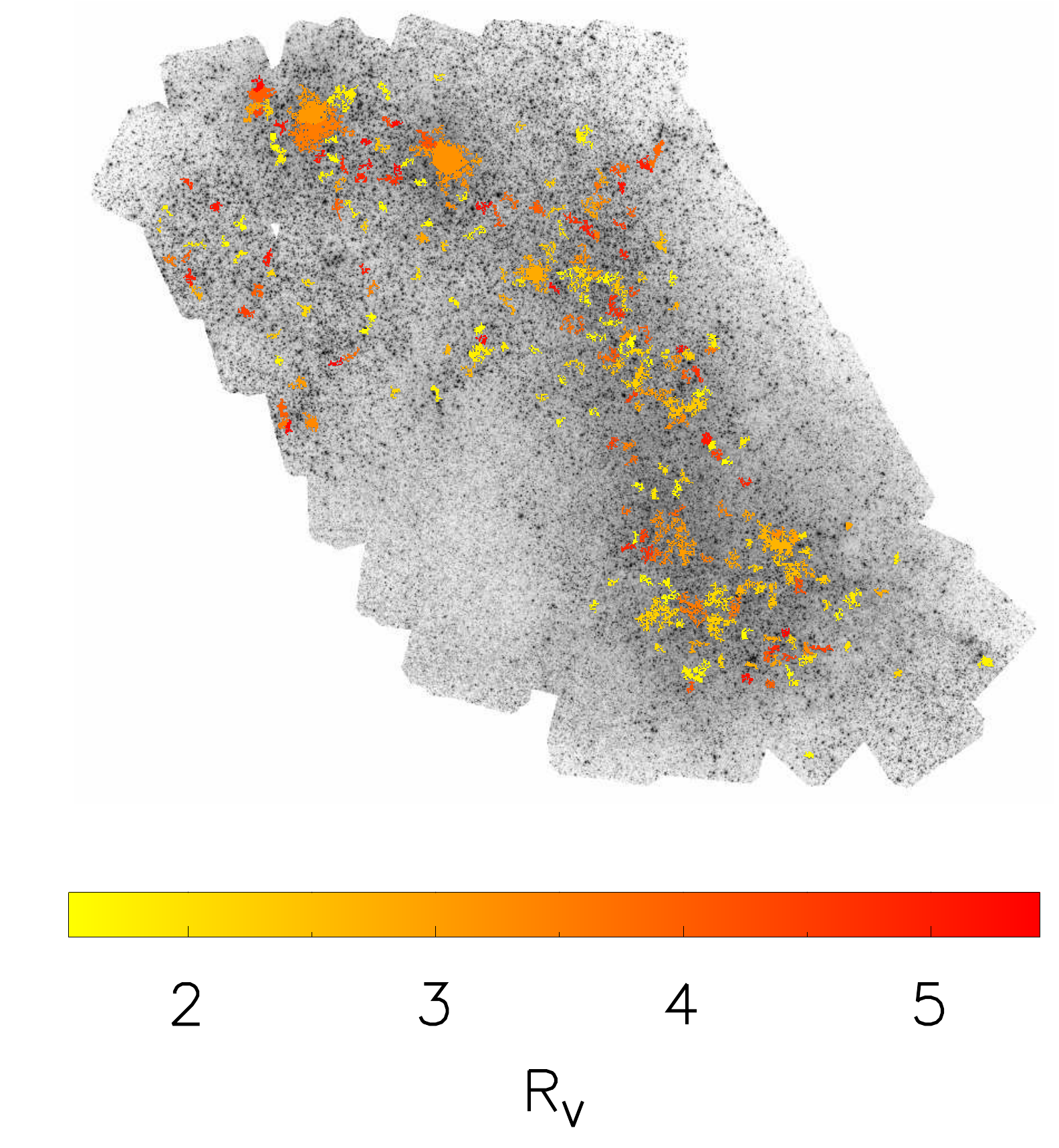}
	\includegraphics[trim = 30mm 30mm 25mm 75mm, clip=true, width=0.3\textwidth]{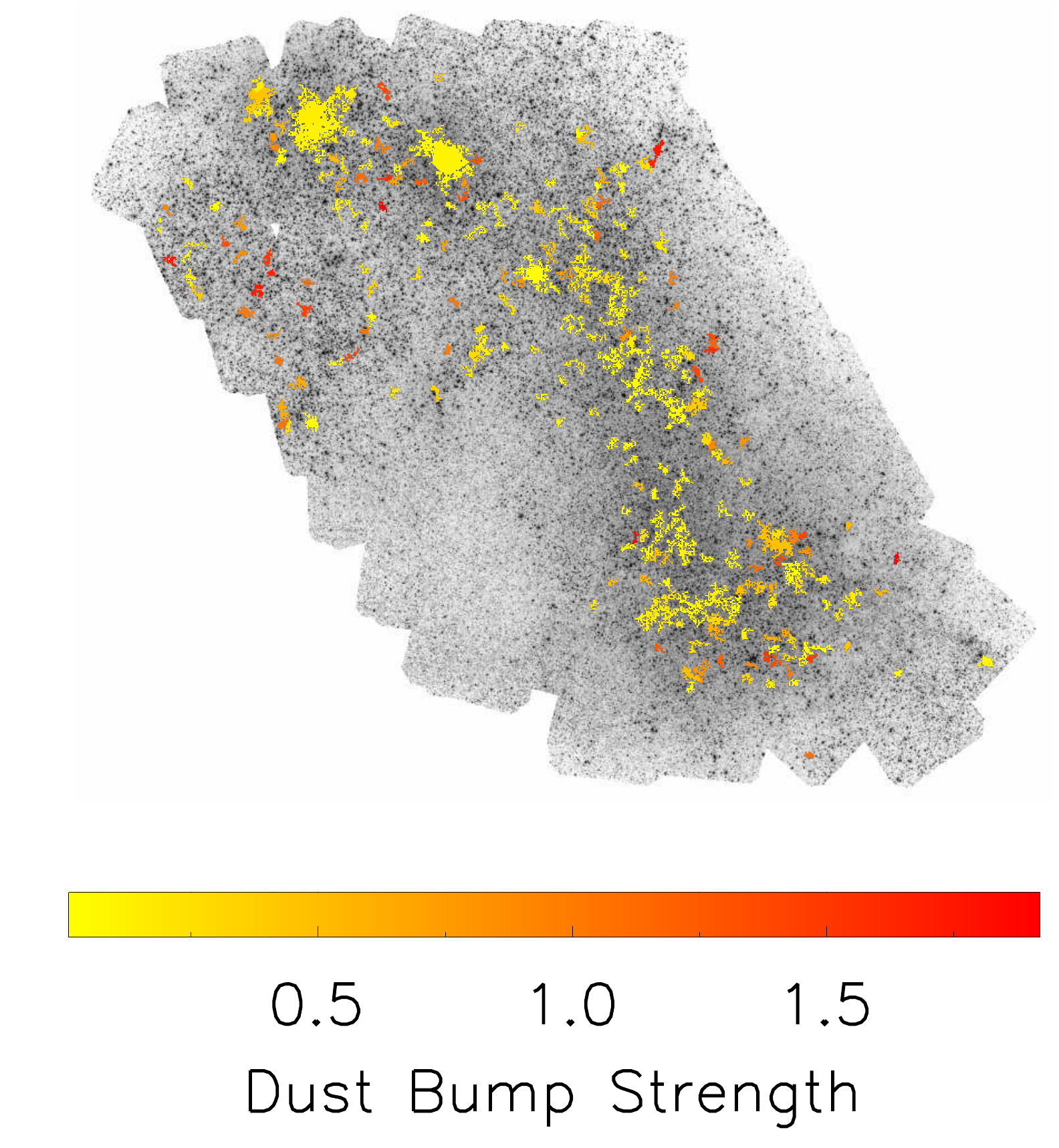}
	\includegraphics[trim = 30mm 30mm 25mm 75mm, clip=true, width=0.3\textwidth]{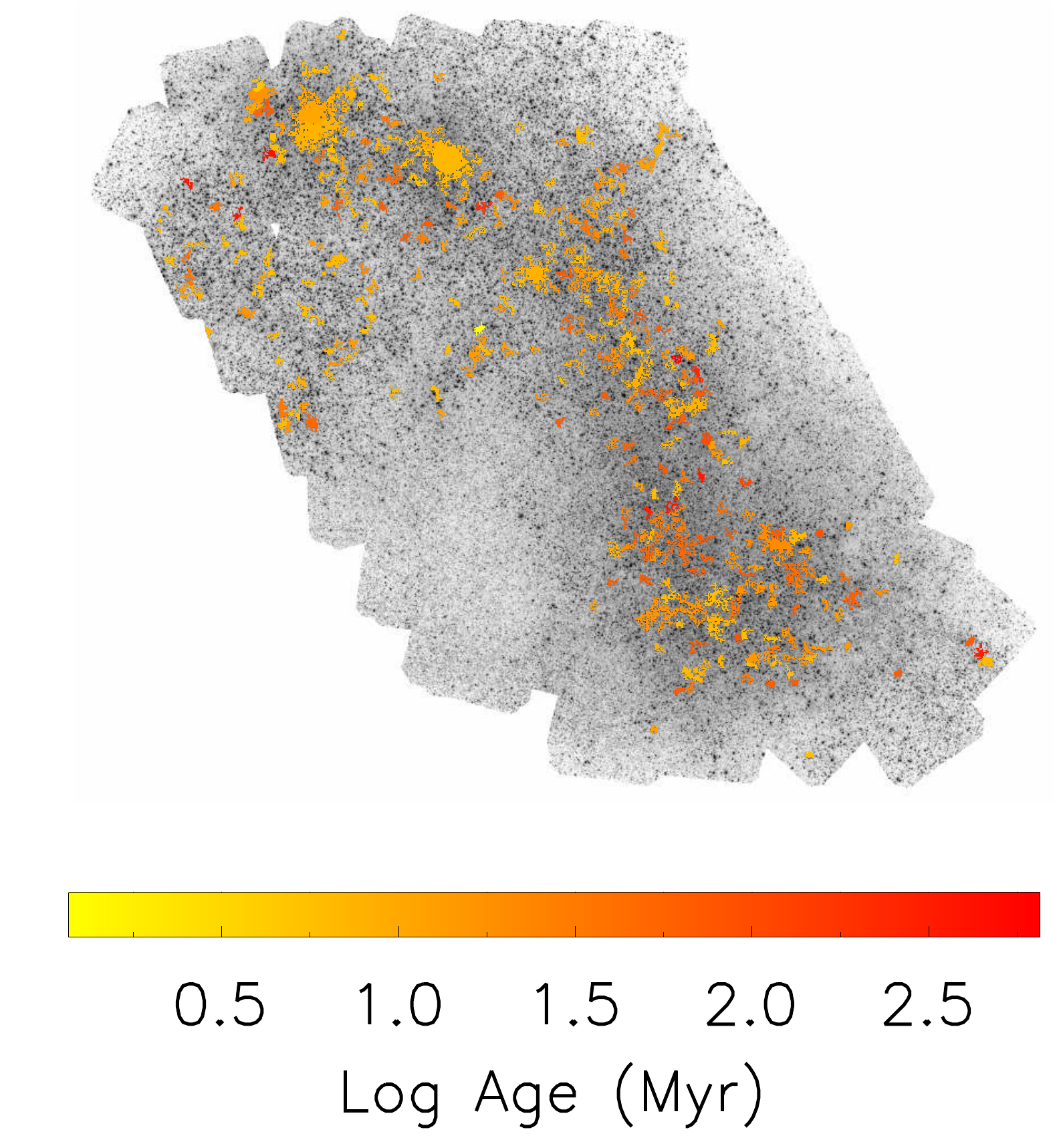}
	\includegraphics[trim = 30mm 30mm 25mm 75mm, clip=true, width=0.3\textwidth]{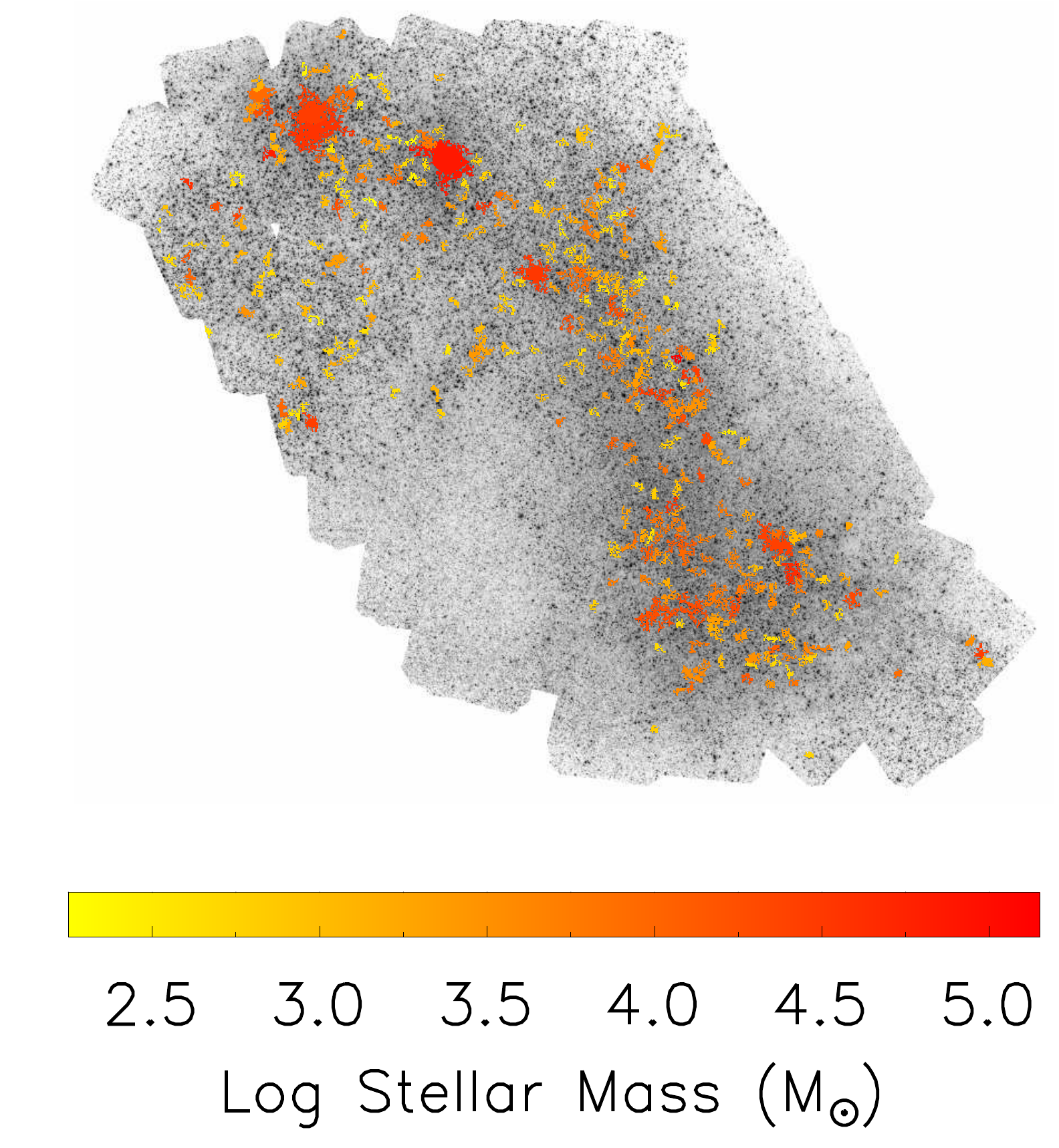}
	\caption{Maps of physical parameters of star-forming regions.}
	\label{fig-sfr_maps}
\end{figure*}


Maps of the physical parameters for the star-forming regions are in Fig.~\ref{fig-sfr_maps}.  The regions only cover 6\% of the UVOT survey area, but it is clear that there is variation on small physical scales.  Where regions appear to be in the same star-forming complex, the ages are typically similar, but the dust curve properties are quite different.  There is no obvious large-scale pattern in any of the physical parameters. 

\begin{figure}
	\centering 
	\includegraphics[trim = 0mm 35mm 0mm 35mm, clip=true, width=0.98\columnwidth]{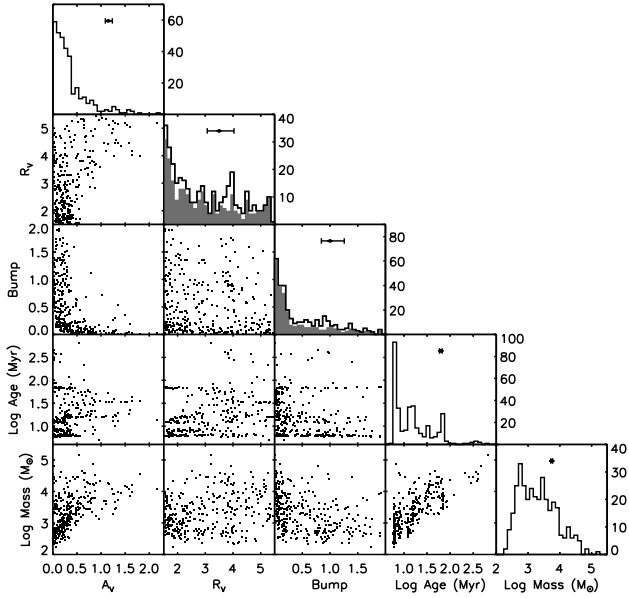}
	\caption{Modeled physical quantities of the star-forming regions plotted against each other.  Shaded histograms for $R_V$ and the bump strength are for regions with $A_V > 0.1$; when there is only a small amount of dust ($A_V \le 0.1$), the extinction curve parameter values cannot be measured.  The mass refers to the stellar mass.  Error bars above each histogram show the median lower and upper uncertainties.  The plot regions with no data points - especially notable in the $A_V$, age, and mass plots - are due to the selection effects discussed in \S\ref{sec-data_red-sfr}.}
	\label{fig-triangle_sfr}
\end{figure}

The parameters for all 338 star-forming regions are plotted against each other in Fig.~\ref{fig-triangle_sfr}, showing to what extent parameters are correlated with each other.   Before discussing the results for each physical parameter, however, it is worth returning to the discussion of selection effects (\S\ref{sec-sf_detect}).  Because these star-forming regions are chosen as UV overdensities, large areas of parameter space are inaccessible.  Therefore, the broad results are not representative of the SMC as a whole.


Most of the star-forming regions have low dust content.  We find that for about a third of the regions, the total dust content is quite low: $33.4 \pm 1.5$\% of the regions have $A_V \le 0.15$ ($E(B-V) = 0.05$ for $R_V = 3.1$).  Approximately half of the regions have $A_V < 0.25$.
When considering the 263 regions with $A_V > 0.1$, and thus have quantifiable $R_V$ and bump values, the dust extinction curves vary considerably.  69\% of the regions have 2175~\AA\ bump strengths consistent with zero at 2$\sigma$, however there is a substantial fraction ($17.1 \pm 1.5$\%) with bumps stronger than the typical MW value (bump $> 1$).  The values for $R_V$ span the whole range from 1.5 to 5.5, indicating a large range of dust curve steepness.

The ages of the star-forming regions are necessarily young.  From the age histogram in Fig.~\ref{fig-triangle_sfr}, there is evidence for enhanced star formation at 6, 15, and 60~Myr ago, \boldchange{though the histogram should not be interpreted as a SFH}.  In addition, given the strong selection against low-mass objects at older ages, drawing any quantitative conclusions about star formation rates is impossible.  \boldchange{A detailed discussion of the SFH of the star-forming regions is deferred to Section~\ref{sec-sfh}.}

\boldchange{In addition, there are many instances of neighboring regions that have dissimilar ages.  \citet{efremov98} and \citet{delafuentemarcos09} find that for the LMC and Milky Way, respectively, there is a positive correlation between the physical separation of pairs of open clusters and the clusters' average age difference.  They interpret this relationship as evidence for star formation that is spatially and temporally hierarchical.  For our star-forming regions, we find no correlation.  The SMC has a line-of-sight depth of $\sim$14~kpc \citep{subramanian12}, which is considerably larger than the physical sizes of the regions, so it is likely that regions with small projected separations are in fact not associated. }

\boldchange{The star-forming regions have a range of stellar masses from 200~\msun\ to $1.5 \times 10^5$~\msun.  We note that at the lower masses, the regions could have stochastically sampled IMFs. While fully quantifying this is beyond the scope of our modeling, it is worth a brief discussion of the possible effects.  \citet{anders13} compare the physical parameters derived from SED modeling of star clusters with and without the assumption of a stochastically sampled IMF.  They find that for clusters with masses less than $10^4$~\msun, not accounting for stochasticity leads to underestimating the modeled mass by 0.2-0.5~dex, underestimating age by 0.1-0.5~dex, and overestimating $E(B-V)$ by 0.05-0.15. However, the 1$\sigma$ uncertainties in these under- and over-estimates are consistent with no offset. }

\begin{figure}
	\centering
	\includegraphics[trim = 5mm 35mm 0mm 35mm, clip=true, width=0.98\columnwidth]{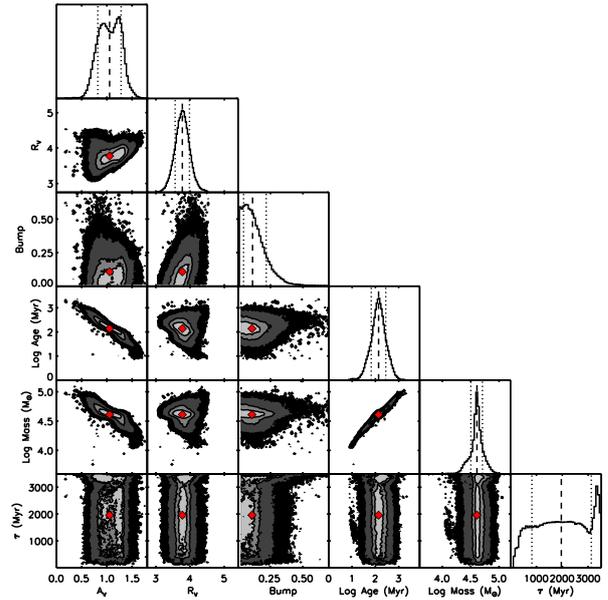}
	\caption{Same as Fig.~\ref{fig-triangle}, but for pixel 499.  As for the star-forming regions, $A_V$, $R_V$, the age, and the stellar mass have significant degeneracies.  The probability distributions for $A_V$ is slightly bimodal, but the uncertainties take this into account. The value of $\tau$ is not strongly constrained.}
	\label{fig-triangle2}
\end{figure}

In Fig.~\ref{fig-triangle2}, we show the uncertainty regions for the physical parameters of an example large pixel.  Similarly to the star-forming region in Fig.~\ref{fig-triangle}, there are degeneracies (correlated uncertainties) between $A_V$, $R_V$, age, and stellar mass, and the 2175~\AA\ bump strength is not degenerate with either of the other parameters.  In addition, it is clear that $\tau$ is not constrained by our data.  For the pixel in Fig.~\ref{fig-triangle2}, as with the bulk of the other pixels, the young ages necessitated by the shape of the SEDs mean that nearly all values for $\tau$ are equally probable.  Finally, the age and mass each have a secondary peak in their probability distributions, though very little probability is contained in these small peaks.  It is worth noting that a typical $\chi^2$ fitting method cannot detect or quantify these types of multi-modal probability distributions; our use of the MCMC modeling assures that the probability distributions are sufficiently well-behaved.

\begin{figure*}
	\centering
	\includegraphics[trim = 30mm 30mm 21mm 75mm, clip=true, width=0.3\textwidth]{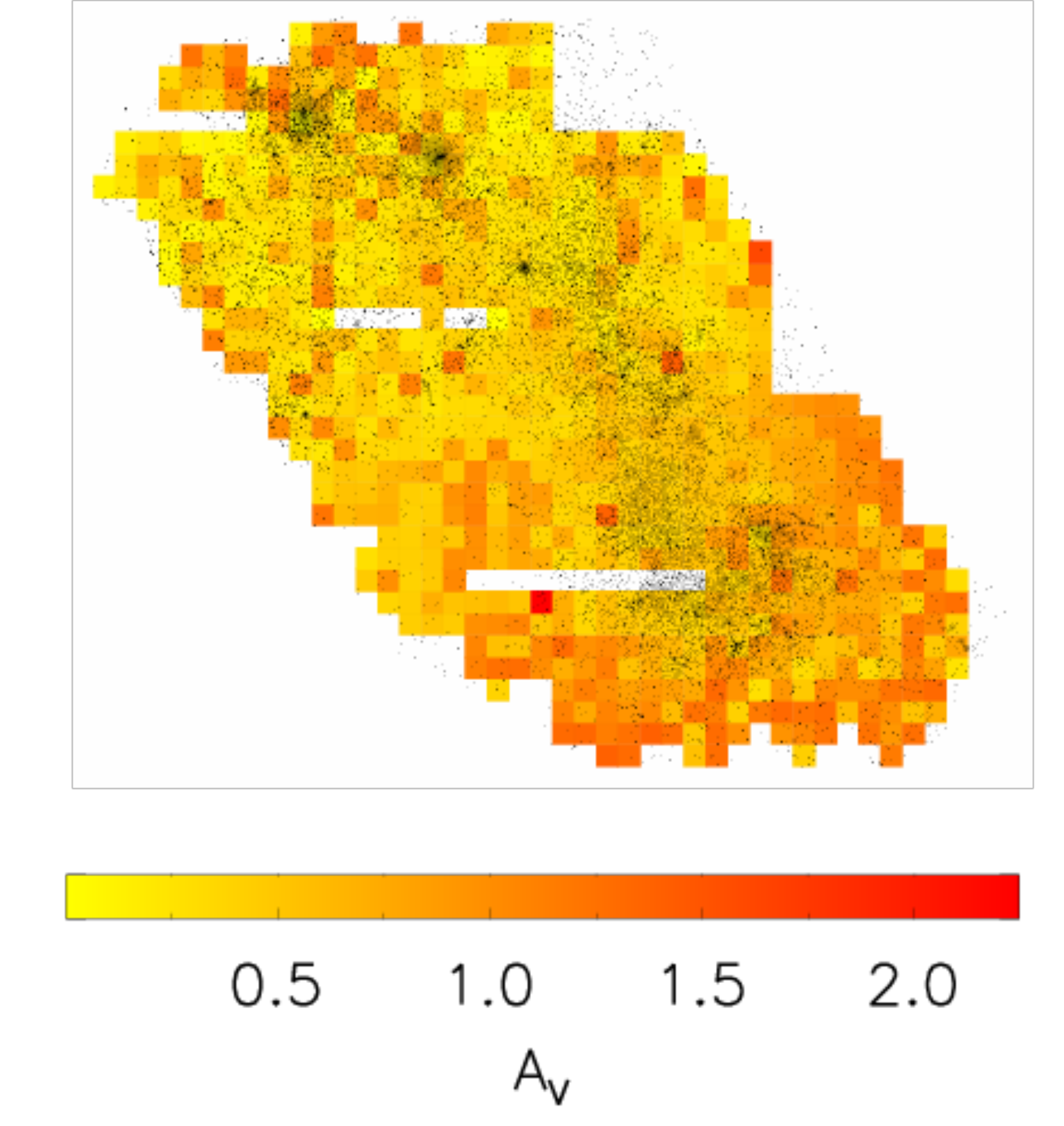}
	\includegraphics[trim = 30mm 30mm 21mm 75mm, clip=true, width=0.3\textwidth]{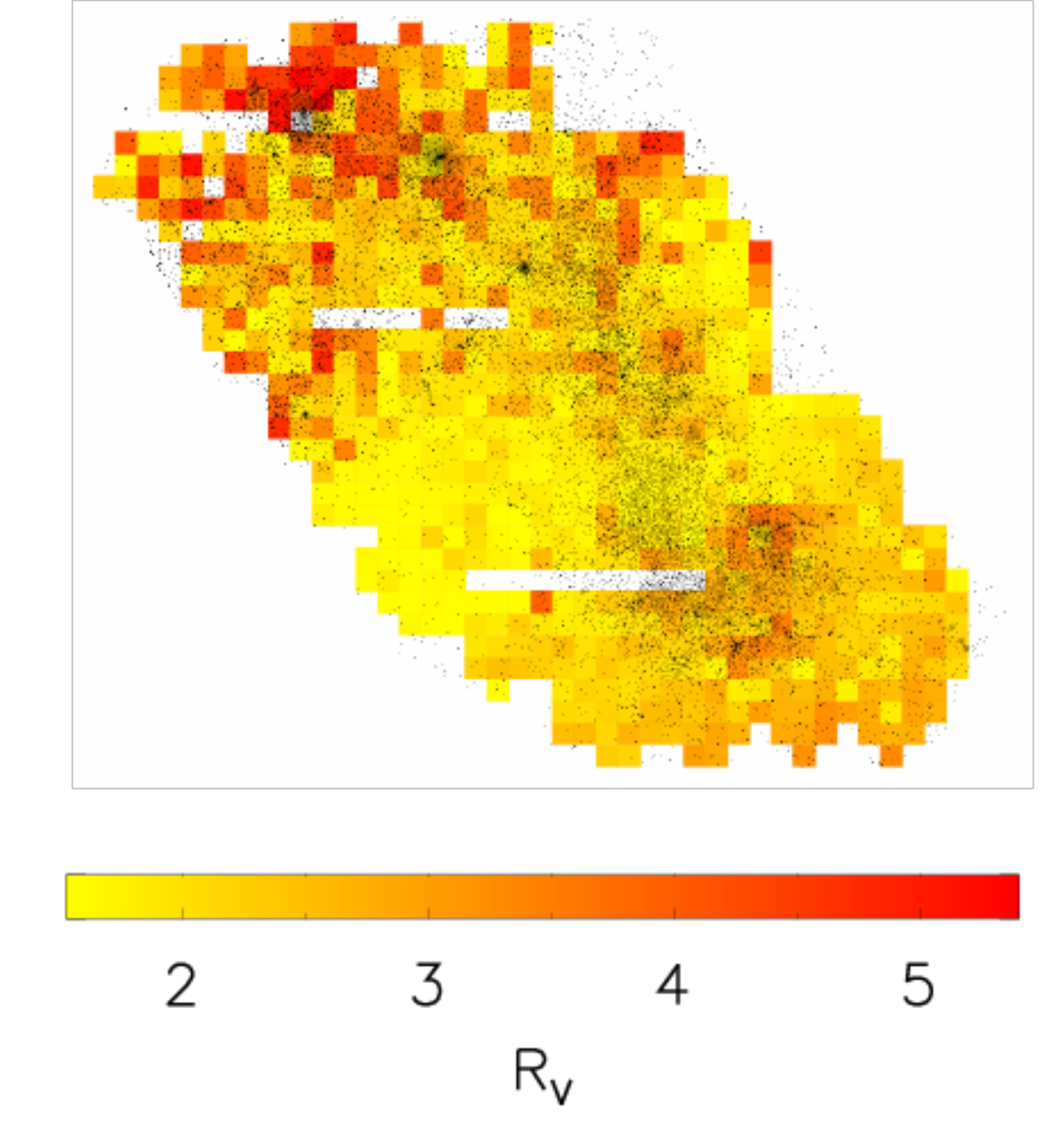}
	\includegraphics[trim = 30mm 30mm 21mm 75mm, clip=true, width=0.3\textwidth]{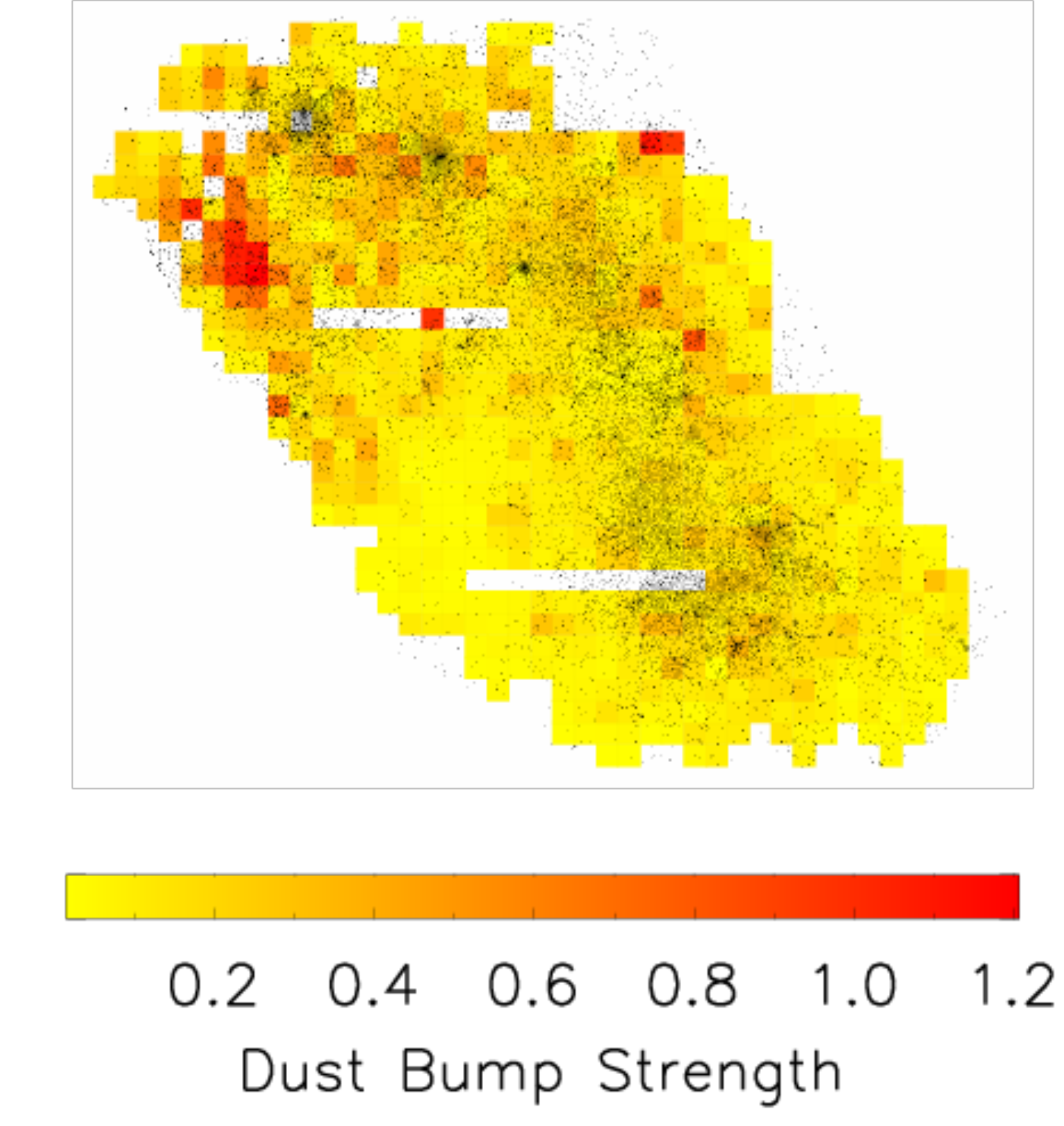}
	\includegraphics[trim = 30mm 30mm 21mm 75mm, clip=true, width=0.3\textwidth]{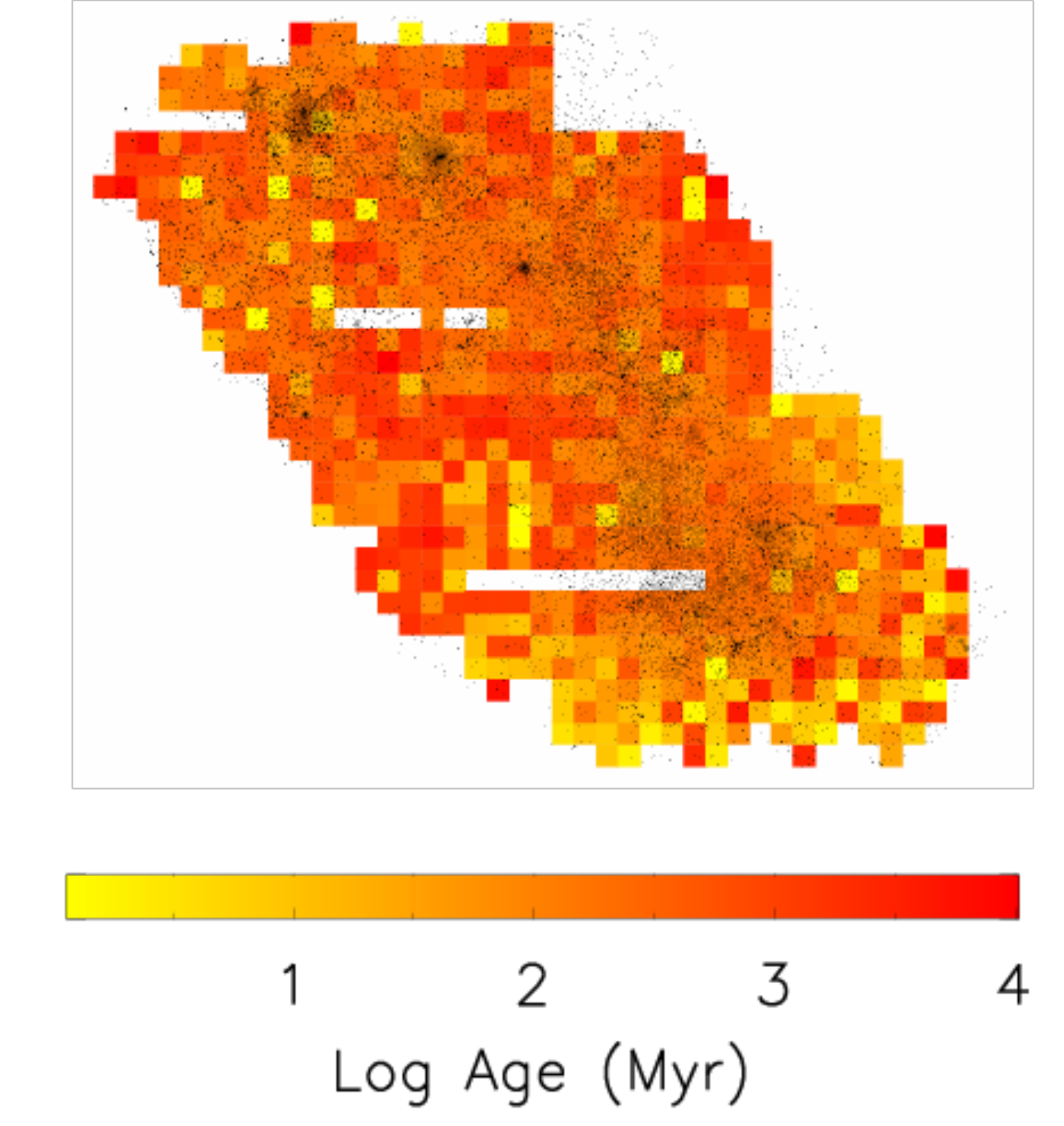}
	\includegraphics[trim = 30mm 30mm 21mm 75mm, clip=true, width=0.3\textwidth]{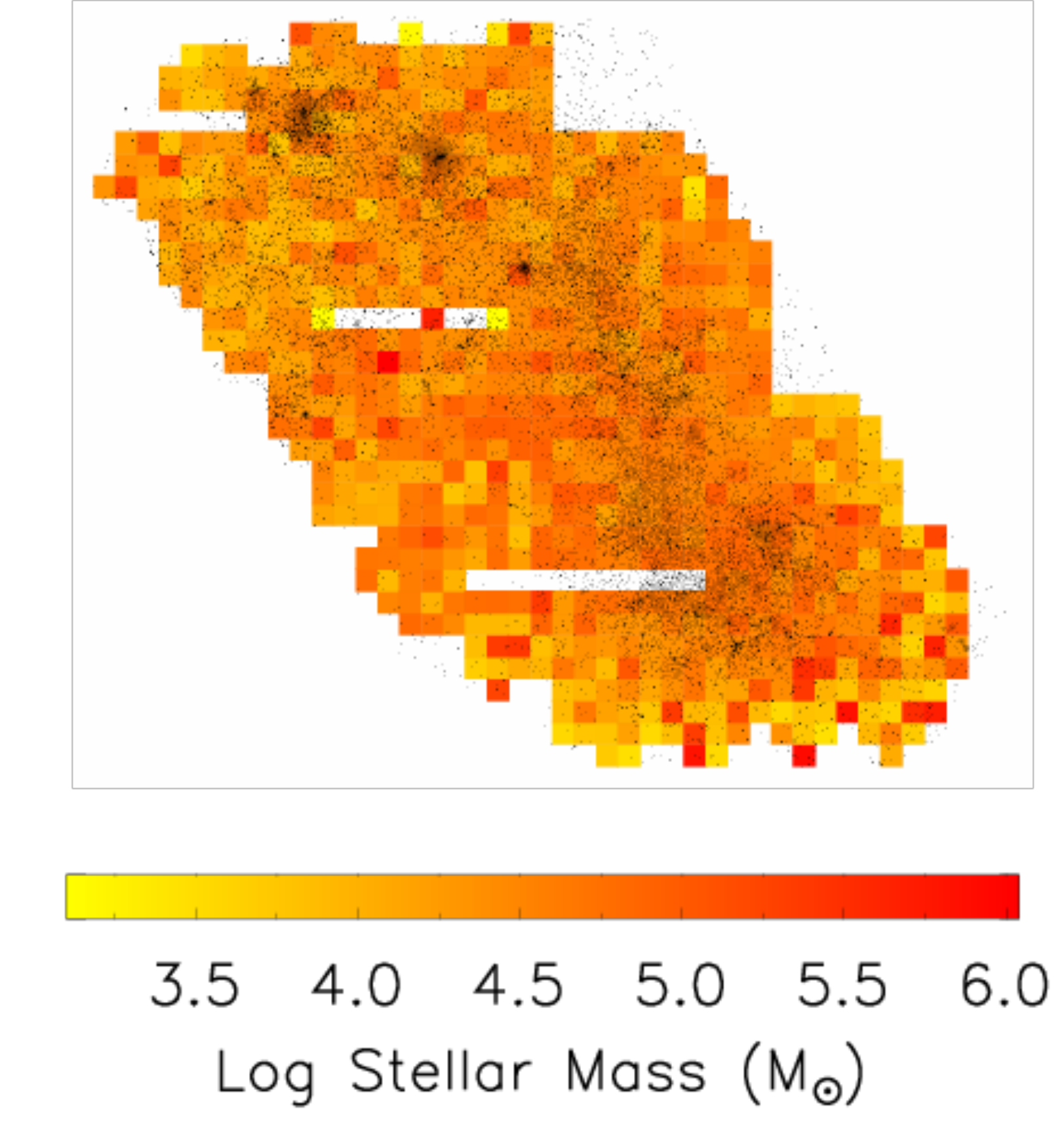}
	\includegraphics[trim = 30mm 30mm 21mm 75mm, clip=true, width=0.3\textwidth]{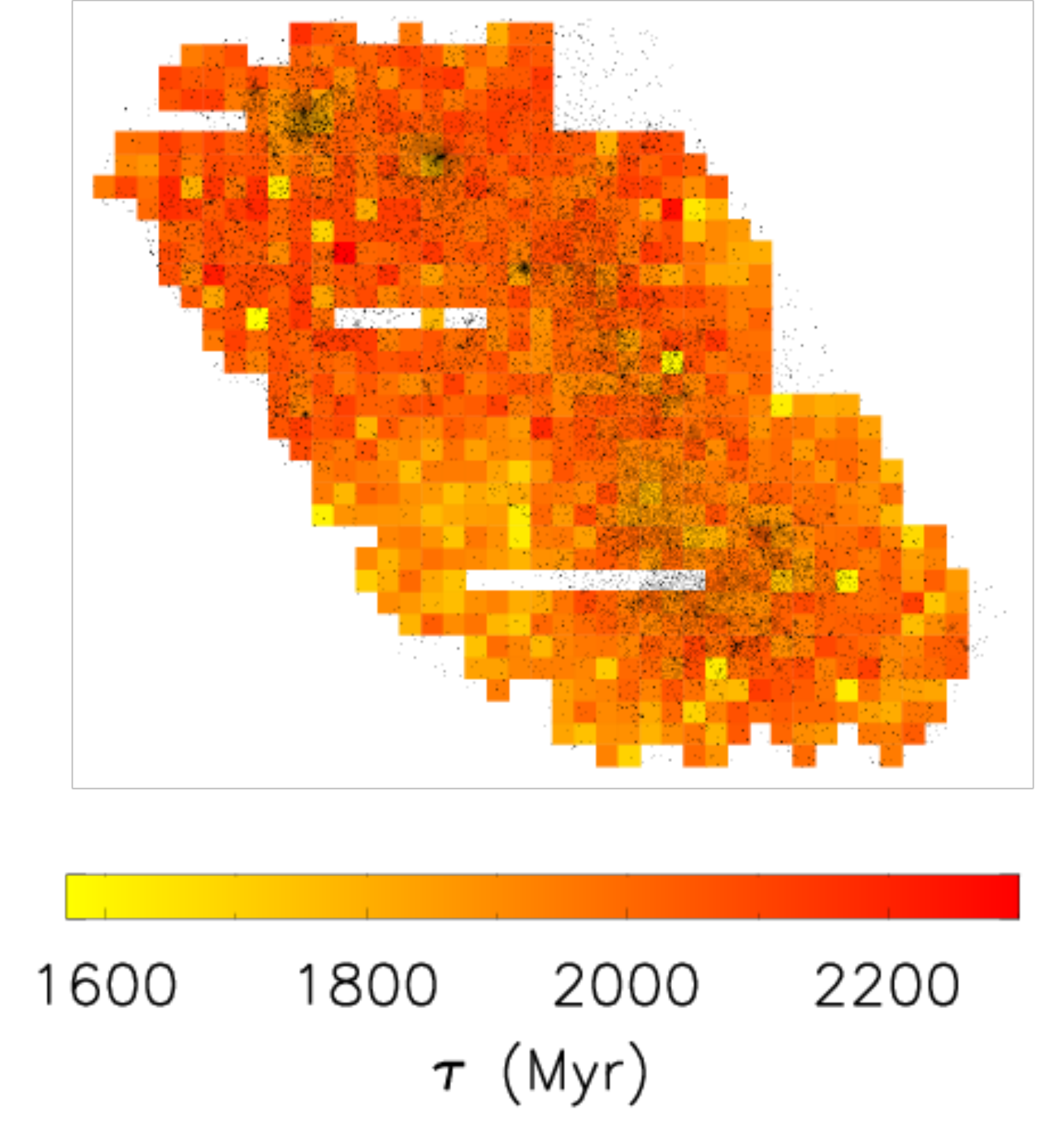}
	\caption{Maps of physical parameters of large pixels.  Overlaid is a greyscale $uvm2$ image for reference.  White areas in the $A_V$, age, stellar mass, and $\tau$ maps were not modeled due to bad pixels in the optical imaging.  Additional white pixels in the $R_V$ and bump images are where $A_V \le 0.1$.}
	\label{fig-pix_maps}
\end{figure*}

The maps of modeled physical parameters for the large pixels are shown in Fig.~\ref{fig-pix_maps}.  With such a large area of the SMC uniformly modeled, one can begin to draw conclusions about the spatial variation of the physical parameters.
The parameters related to dust ($A_V$, $R_V$, bump strength) have immediately apparent trends with position.  $A_V$ tends to be lower to the northeast and higher in the southwest, with relatively lower values along the UV-bright bar.  Similarly, the 2175~\AA\ bump strength has a northeast-southwest gradient, with little to no measurable bump in the southwest and a stronger bump in the northeast.  The values for $R_V$ are generally low and do not appear to have a correlation with position.
The ages along the UV-bright bar are slightly younger than the surrounding areas, though this is difficult to see in the map in Figure~\ref{fig-pix_maps} because of the scaling.  
The values for $\tau$ are not well constrained by the data and have large uncertainties, so any large scale trends should not be overinterpreted.


\begin{figure}
	\centering 
	\includegraphics[trim = 0mm 35mm 0mm 25mm, clip=true, width=0.98\columnwidth]{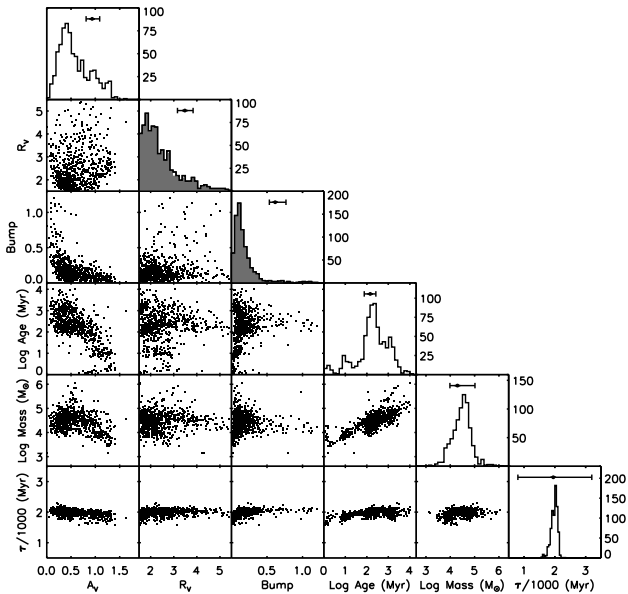}
	\caption{Same as Fig.~\ref{fig-triangle_sfr}, but for the large pixels.  }
	\label{fig-triangle_pix}
\end{figure}

We plot each of the large $200''$ (58~pc) pixel modeled parameters against each other in Fig.~\ref{fig-triangle_pix}.  The figure also has histograms of the physical parameters.  As compared to the star-forming regions, there are not strong selection effects that eliminate areas of parameters space.  \boldchange{However, each pixel is composed of a variety of stellar populations with different formation histories and dust properties, which has the effect making it impossible to assign one precise value for each physical property.  For instance, towards areas of recent star formation, there are large variations in the dust over small physical scales, and the coarseness of the pixels means that we cannot capture any differential extinction.  This is reflected in the uncertainties, which are significantly larger than those for the star-forming regions.  One can view the best-fit physical parameters for each pixel as some weighted average of the constituent populations.}


The dust parameters have fairly tight distributions.  The total dust content $A_V$ is centred around 0.4~mag, with 55\% of pixels between 0.2 and 0.6~mag.  For pixels with $A_V > 0.1$, the dust extinction curve is typically steep: $62 \pm 1$\% have $R_V < 2.5$, with the remainder between 2.5 and 5.5.  
The 2175~\AA\ bump distribution has a median of 0.14, and the largest measured bump is 1.21.
The age distribution of the large pixels has a large peak at 150~Myr, with smaller peaks at 1.5~Myr, 10~Myr, and 1~Gyr.  We note that age and star formation timescale $\tau$ are mathematically degenerate, and $\tau$ is not strongly constrained by our observations.  Likewise, the uncertainties for the stellar mass are large because of the large uncertainty of $\tau$.

It is interesting to compare the distributions of modeled physical parameters for the star-forming regions and large pixels.  Both have a total dust $A_V$ concentrated below $A_V = 0.5$, with tails extending to larger $A_V$.  The distributions of dust curve parameters ($R_V$ and bump strength) are strikingly different: they are much flatter for the star-forming regions than for the pixels.  However, the $R_V$ distributions both peak at low $R_V$, and the bump strengths for both tend to be lower than 0.5.


The age distributions are also somewhat different.  While nearly all of the star-forming regions have ages younger than 100~Myr, the large pixels are primarily centred around 150~Myr.  Both have several distinct age peaks, including an overlapping peak at $\sim$10~Myr.  As already discussed, the large pixels necessarily average over several populations, so the age resolution is not as high.  In addition, since we assume different star formation histories for the regions and pixels, the ages have different meanings.  This is addressed further in Section~\ref{sec-sfh}.

\input{table6.tex}


\section{Discussion} \label{sec-disc}

\subsection{Implications for Dust Composition}

Here we compare our results to those derived at other wavelengths.  Our analysis primarily focuses on the large pixel modeling results because they are not subject to selection effects that could bias the results.  In addition, they span the whole survey area, so our conclusions are valid for the whole SMC.

\begin{figure}
	\centering
	\includegraphics[trim = 30mm 30mm 25mm 75mm, clip=true, width=0.9\columnwidth]{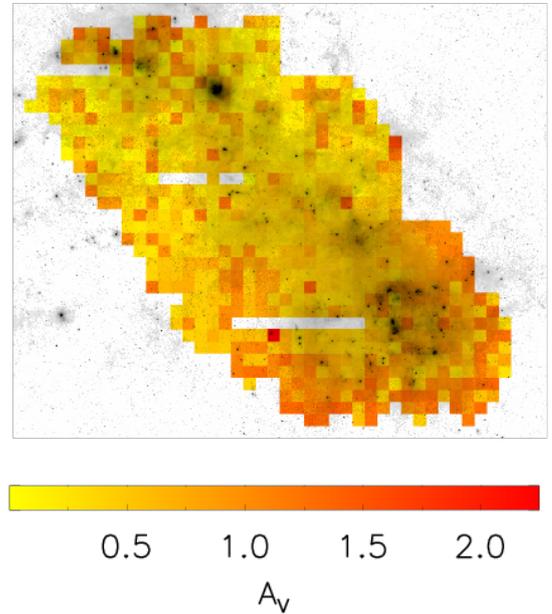}
	\caption{Map of $A_V$ overlaid with the MIPS 24~$\mu$m image \citep{gordon11}.  While the lowest values of $A_V$ occur along the UV-bright areas of star formation (Fig.~\ref{fig-pix_maps}), they are offset from the highest 24~$\mu$m emission.}
	\label{fig-av24}
\end{figure}

First, we compare $A_V$ to the 24~$\mu$m imaging from SAGE-SMC \citep{gordon11} in Fig.~\ref{fig-av24}.  Since 24~$\mu$m emission traces dust, one would expect that the highest $A_V$ would correspond to bright 24~$\mu$m regions.  We find that there is indeed a spatial correlation on the largest scales. \boldchange{This is consistent with the discussion in Section~\ref{sec-results} that the large pixels cannot measure differential extinction; our $A_V$ values necessarily trace the more diffuse dust content of each pixel rather than any small-scale clumpy components}.  In the southwest, there is a large ring-shaped feature in the 24~$\mu$m map that overlaps with the area of highest $A_V$.  To the east of this feature is a region of lower $A_V$, which corresponds to a peak of UV emission (top-left panel of Fig.~\ref{fig-pix_maps}).  This implies that the recent star formation has blown away much of the dust, which is consistent with the age map, which has a comparatively older population at that location.

\boldchange{We also find that for the large pixels, $A_V$ is broadly correlated with the ratio of the 24~$\mu$m and UV fluxes: for pixels with larger $A_V$, there is a higher 24~$\mu$m flux compared to the fluxes at either $uvw2$, $uvm2$, or $uvw1$.  This means the presence of dust is being captured by the suppression of UV light. While it would be optimal to include the full near- and mid-IR SED to trace the emission of dust as part of our modeling, we can safely defer it to future work.}

Another location that is interesting to consider is the star-forming region NGC~346, which is the bright concentrated UV source on the northern end of the SMC in Fig.~\ref{fig-pretty}.  NGC~346 is also extremely bright in the 24~$\mu$m image, indicating a large amount of dust.  Since the region is so young \citep{cignoni11}, it is bright in UV but hasn't had time to blow away its surrounding dust.  In the large pixel $A_V$ map, our modeling suggests a low $A_V$ of $\sim$0.4, but NGC~346 is modeled as a single star-forming region (Fig.~\ref{fig-sf_map}), for which we measure a much higher $A_V = 0.83$.

A map of $A_V$ values has been derived by \citet[][\citetalias{zaritsky02}]{zaritsky02} using optical broadband photometry to model the individual stars' atmospheres and foreground dust attenuation.  One map each was created for the hot stars (12000~K $<T_\text{eff} < $ 45000~K) and cool stars (5500~K $< T_\text{eff} <$ 6500~K) with pixel scales of $60''$.  The dust attenuation is typically higher for the hotter stars, which \citetalias{zaritsky02} attributes to dust surrounding the hot stars on small physical scales.

An explicit comparison of our map to those of \citetalias{zaritsky02} is not actually informative.  First, our UV observations make us sensitive to the hotter UV-bright stellar populations.  However, our results are for combinations of stars of different histories and temperatures, so it is impossible to make a direct comparison to either the hot or cool stars in \citetalias{zaritsky02}.  Second, the uncertainties in $A_V$ for both for our method and that of \citetalias{zaritsky02} are of order 0.1~mag, and the majority of measured $A_V$ values are in a range of several tenths of magnitudes.  Since the errors are of similar scale to the range of measured values, the scatter in a comparison of our $A_V$ values would be so large as to obscure any relationship.

It is worth noting that the dust properties one uses for a particular purpose is very much dependent upon the corresponding analysis.  Our grid of models assume a stellar population with a plane of foreground dust, but the SMC clearly has three-dimensional structure \citep{mathewson86,welch87}.  Therefore, one should use our $A_V$ and dust curve values with care.



\begin{figure*}
	\centering
	\includegraphics[trim = 30mm 30mm 20mm 75mm, clip=true, width=0.45\textwidth]{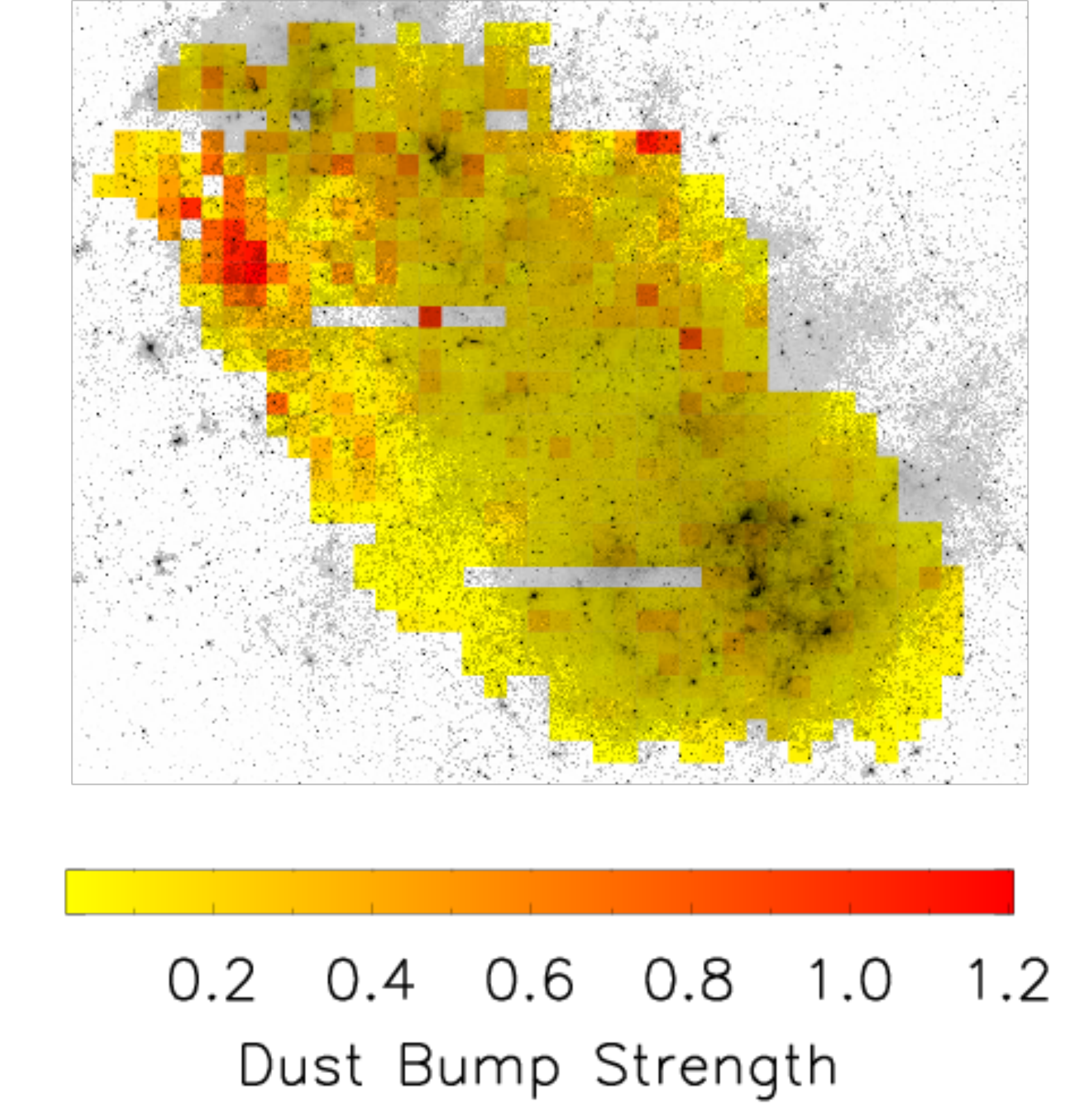}
	\ \ \ \ 
	\includegraphics[trim = 35mm 95mm 35mm 30mm, clip=true, width=0.45\textwidth]{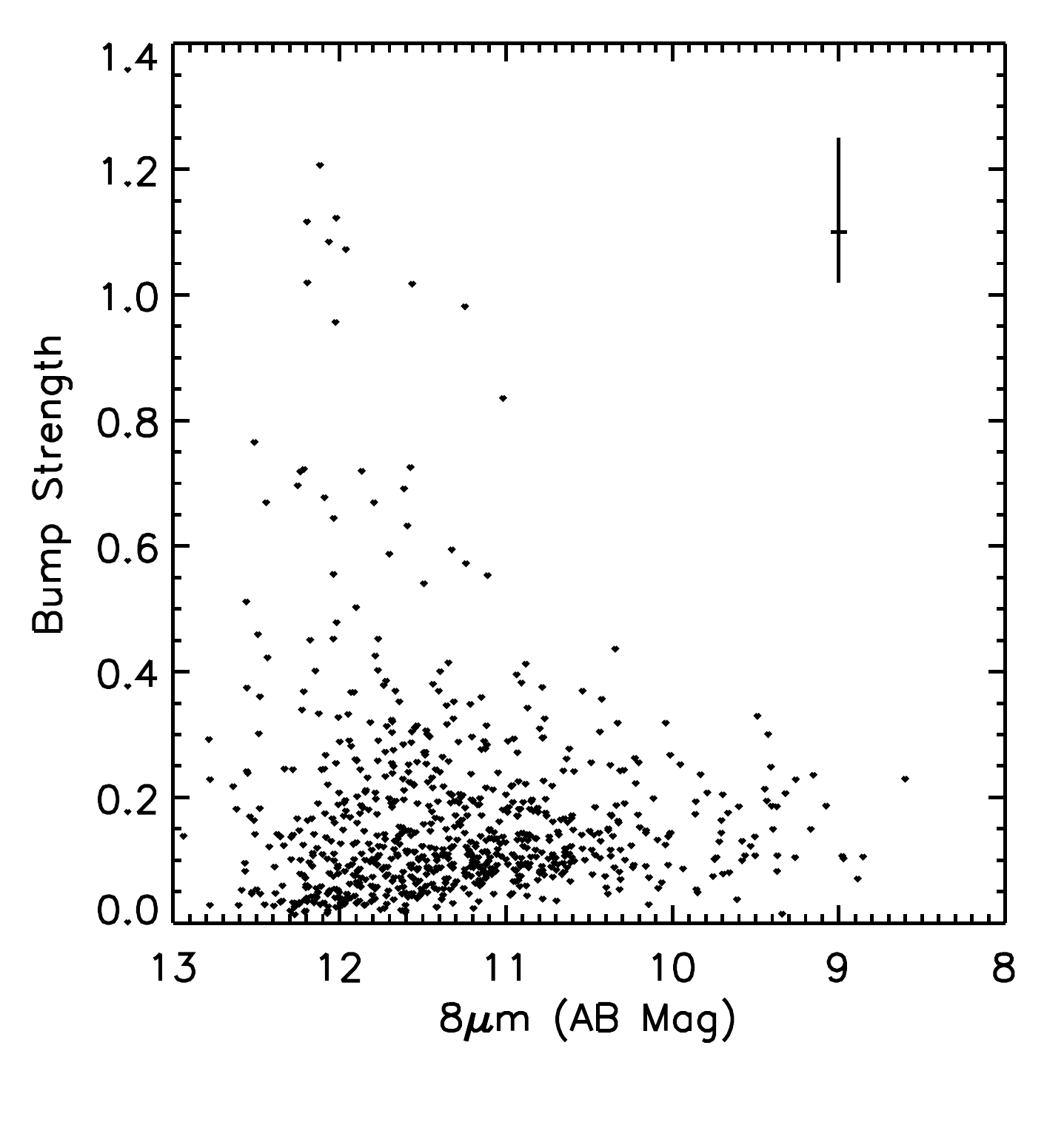}
	\caption{Comparison between the 2175~\AA\ bump strength and IRAC 8~$\mu$m emission, which traces PAH emission.
	\textit{Left:} Map of the bump strength overlaid with the 8~$\mu$m image \citep{gordon11}.  
	\textit{Right:} For each large pixel, the bump strength plotted against the 8~$\mu$m magnitude, with the median errors shown at the top right.
	Both panels show that the largest bump strengths are associated with lower 8~$\mu$m emission, and high 8~$\mu$m emission only corresponds to low bump strength.}
	\label{fig-bump8}
\end{figure*}

In Fig.~\ref{fig-bump8}, we show the SAGE-SMC 8~$\mu$m image overlaid on the 2175~\AA\ bump strength imaging.  It has been suggested that the bump is caused by absorption by PAHs \citep{li97}, which also have emission bands in the 8~$\mu$m band \citep{allamandola89}.  However, we do not find evidence for this correlation.  In the image, the areas with the largest bump strength (primarily to the northeast) are located where there is less 8~$\mu$m emission.  This is also demonstrated in the plot of bump strength and 8~$\mu$m brightness in Fig.~\ref{fig-bump8}.  While there is no evidence of a correlation between the bump strength and 8~$\mu$m emission, the plot is lacking in points with high bump strength and high 8~$\mu$m flux.  The pixels with the highest 8~$\mu$m flux have low bump strengths, and the pixels with the largest measured bumps have lower 8~$\mu$m flux.  A third of the pixels ($35 \pm 1$\%) have both a small dust bump (below 0.3) and low 8~$\mu$m flux (fainter than 11~mag).

\begin{figure}
	\centering
	\includegraphics[trim = 35mm 105mm 35mm 35mm, clip=true, width=0.9\columnwidth]{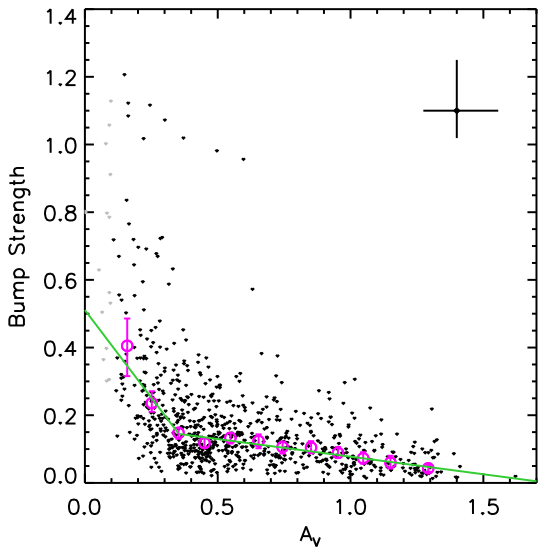}
	\caption{Correlation between bump strength and $A_V$ for the large pixels, with the median uncertainty shown at the top right.  Magenta circles are medians within twelve $A_V$ bins, with error bars marking the uncertainty of the median.  The green lines are the best-fitting linear models to the low-$A_V$ and high-$A_V$ magenta medians.  Grey points have $A_V < 0.1$ and were not used in the fitting process.}
	\label{fig-bump_av}
\end{figure}

In Fig.~\ref{fig-bump_av}, we plot the 2175~\AA\ bump strength as a function of $A_V$.  Only considering the points with $A_V > 0.1$, we find that two linear fits are necessary to describe the data: one for the steep relationship at low $A_V$ and one for the shallower relationship above $A_V \approx 0.4$.
\begin{equation}
B =
\begin{cases}
(0.51 \pm 0.10) + (-1.04 \pm 0.32) A_V & 0.1 < A_V < 0.35 \\
(0.18 \pm 0.01) + (-0.10 \pm 0.02) A_V & A_V \geq 0.35 \\
\end{cases}
\end{equation}

The four \citetalias{gordon03} stars in the SMC bar have $A_V$ values between 0.35 and 0.68, and with their negligible 2175~\AA\ bumps, they are within the scatter of the points in Fig.~\ref{fig-bump_av}.  
Previous work has suggested that stronger dust bumps are associated with larger reddening at higher redshifts of $1 < z < 2.5$ \citep[e.g.,][]{noll05,noll07}, which may be due to metallicity effects.  However, it is unclear what the underlying physical reason is for these correlations.


\subsection{Comparison to \citet{gordon03}}

\begin{figure*}
	\centering
	\includegraphics[trim = 30mm 30mm 20mm 75mm, clip=true, width=0.4\textwidth]{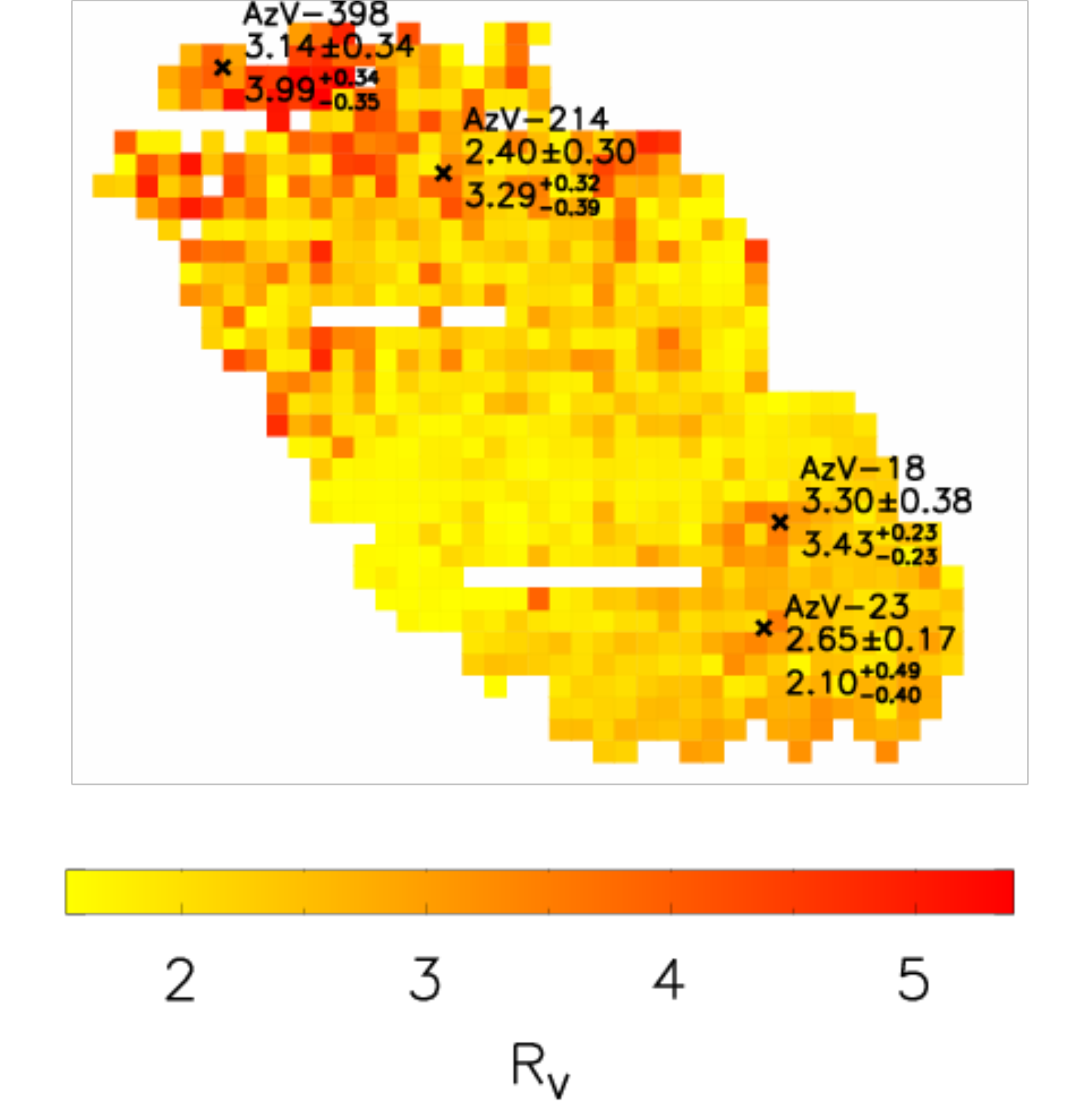}
	\includegraphics[trim = 30mm 30mm 20mm 75mm, clip=true, width=0.4\textwidth]{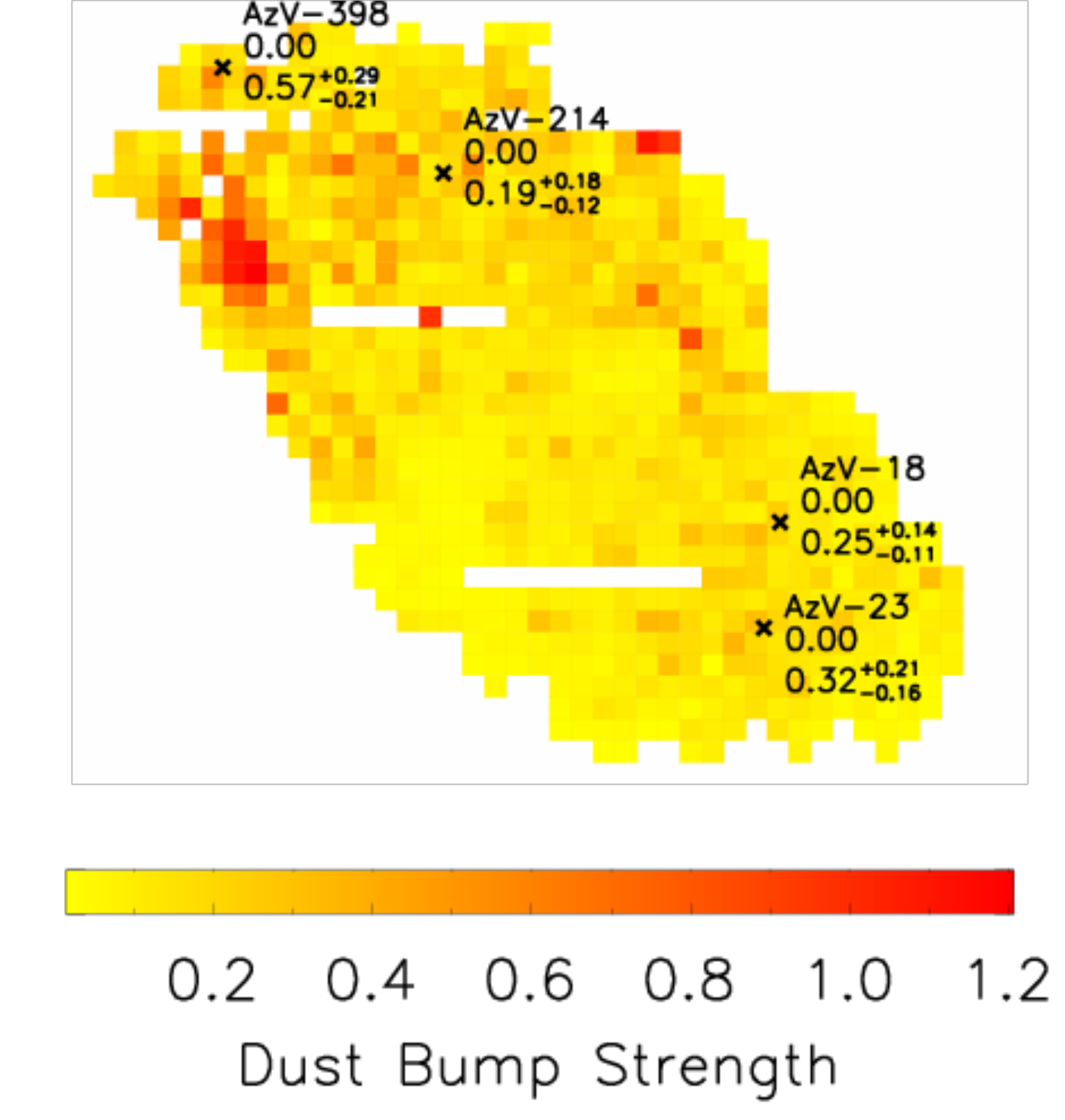}
	\includegraphics[trim = 15mm 80mm 35mm 45mm, clip=true, width=0.5\textwidth]{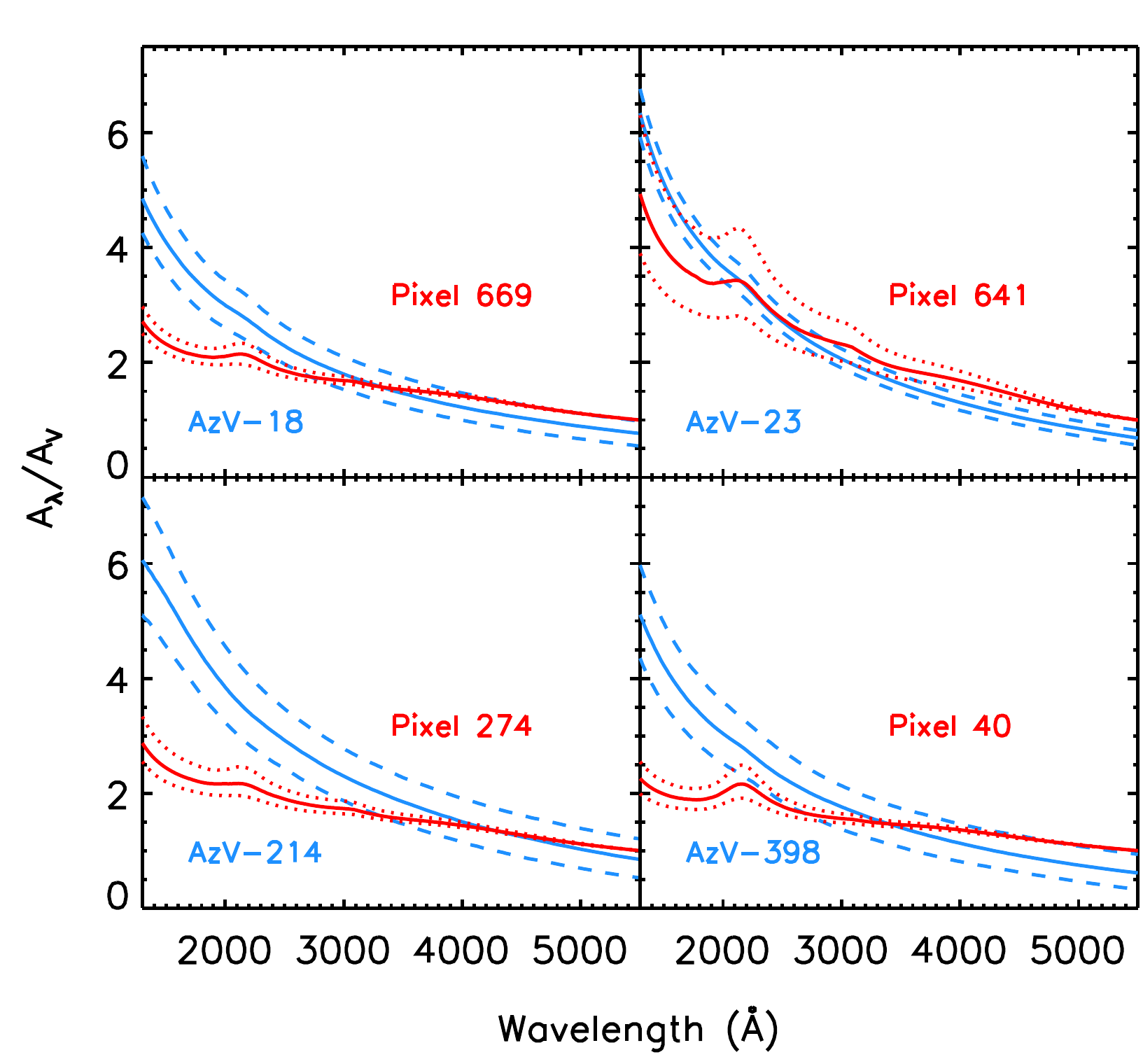}
	\caption{Comparison between \citetalias{gordon03} extinction curve measurements and the large pixel parameters at the same location.  \textit{Top:} Maps of $R_V$ and bump strength with the four \citetalias{gordon03} stars marked.  The labels show the name of the star, the \citetalias{gordon03} measurement, and our modeled value.  \textit{Bottom:} Dust curves for the stars and modeled pixels, with $1\sigma$ uncertainties.}
	\label{fig-g03}
\end{figure*}

\citetalias{gordon03} is the classic reference for the SMC dust extinction law.  At the top of Fig.~\ref{fig-g03}, we show the locations of the four ``SMC Bar" stars (the ``SMC Wing" star is not in our observed area) on the $R_V$ and bump strength maps.  Each star is labelled with its name, \citetalias{gordon03} $R_V$ or bump strength, and our measured $R_V$ or bump strength.  Below these maps are the corresponding extinction curves for each star.

When comparing our values for $R_V$, it is important to note that \citetalias{gordon03} quantifies the extinction curves following \citet{fitzpatrick90}, which has seven shape parameters, and $R_V$ is used in a mathematically different manner than by our adopted \citet{cardelli89} formalism.
From the full dust curves in Fig.~\ref{fig-g03}, the slope for star AzV-23 are almost identical to our modeled slopes.  The curves for the other three stars (AzV-18, AzV-214, and AzV-398) are somewhat steeper, but they are statistically indistinguishable for wavelengths longward of $\sim$2500~\AA.  The addition of far-UV (FUV) imaging data for the SMC would improve the ability of our modeling to constrain the slope at shorter wavelengths.

For the 2175~\AA\ dust bump strength, we measure a significant bump at the positions of three of the four stars, whereas \citetalias{gordon03} assume no bump.  The bump strength for our large pixel overlapping the star AzV-214 is consistent with zero at $2\sigma$.  
For the other three stars, our measured bump strengths are considerably larger (0.25 to 0.57 of the Milky Way bump).  Star AzV-398 is on the corner of a pixel, and the neighboring three pixels have comparatively smaller bump strengths of 0.21, 0.15, and 0.22.  It is known that the dust extinction curve properties can vary on small physical scales \citep[e.g.,][]{demarchi16}, so it is reasonable that the \citetalias{gordon03} extinction curves do not perfectly match those of the corresponding pixels.

\subsection{Recent Star Formation History} \label{sec-sfh}

We have two options for deriving the recent SFH of the SMC: using the modeling results from either the star-forming regions or the large pixels.  Both sets of results have advantages and disadvantages for calculating a SFH.  The star-forming regions are, by definition, areas with bright UV emission indicating recent star formation.  However, they are subject to the biases discussed in Section~\ref{sec-data_red-sfr}.  Clusters of star formation that are older or more dust-obscured are more strongly selected against and are therefore less likely to be counted.  The large pixels, on the other hand, do include all UV light emission.  But much of the SMC is dominated by older stellar populations, and since we model each pixel with a single exponential SFH, it is impossible to accurately separate the youngest populations from their surrounding older populations.

\begin{figure}
	\centering
	\includegraphics[trim = 25mm 55mm 20mm 60mm, clip=true, width=0.9\columnwidth, page=2]{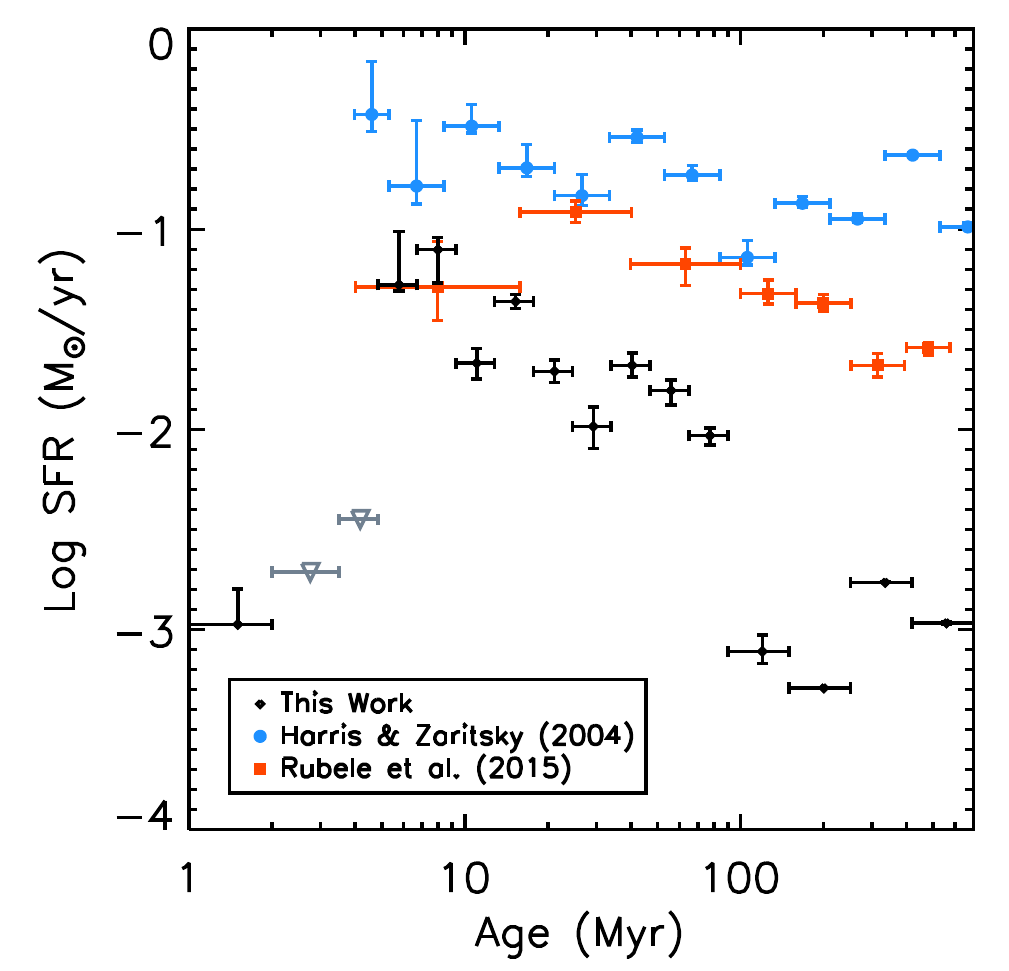}
	\caption{Recent SFH of the SMC as derived from the star-forming regions.  Grey triangles are $3\sigma$ upper limits.  Note that these measurements are subject to the biases discussed in Section~\ref{sec-data_red-sfr} (most notably the lower detection threshold with increasing age), so these are necessarily lower limits.  Blue and red points are the SFHs from \citet{harris04} and \citet{rubele15}, respectively. }
	\label{fig-sfh_sfr}
\end{figure}

\begin{figure}
	\centering
	\includegraphics[trim = 25mm 55mm 20mm 60mm, clip=true, width=0.9\columnwidth, page=1]{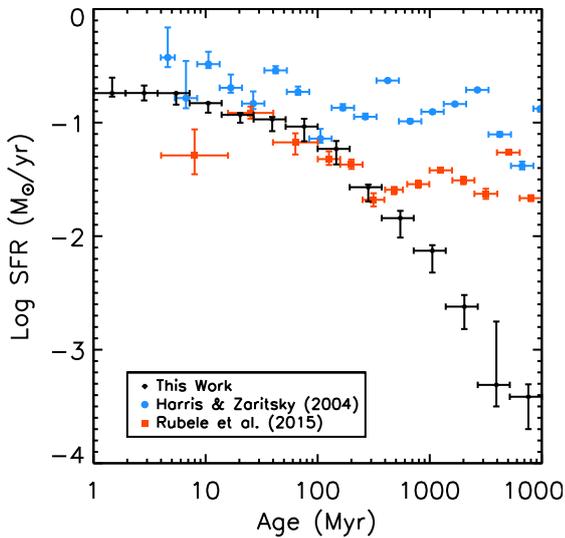}
	\caption{Recent star formation history of the SMC as derived from the large pixels.  Blue and red points are the SFHs from \citet{harris04} and \citet{rubele15}, respectively.}
	\label{fig-sfh_pix}
\end{figure}

Given these considerations, we derive the recent SFH for both the star-forming regions (Fig.~\ref{fig-sfh_sfr}) and large pixels (Fig.~\ref{fig-sfh_pix}) and compare them. 
We calculate the SFH of the SMC using the ages and total processed mass of each star-forming region or pixel.  The processed mass is the total amount of mass that has been turned into stellar material (including stellar remnants).  We derive uncertainties using a Monte Carlo approach, in which we vary the values of mass, age, and $\tau$ within their uncertainties, calculate the SFH, and repeat several hundred times.

Unsurprisingly, the SFHs found from the star-forming regions and large pixels are quite different: the former is somewhat stochastic, whereas the latter is smoothly increasing to the present time.  The star-forming regions represent the brightest  concentrations of UV light, so their distinct ages and bursty star formation lend themselves to a more varied SFH.  Even with this variation, there is a clear increase in the SFR between about 100 and 8~Myr ago, but this is likely an artefact of the decreasing sensitivity with increasing age (\S\ref{sec-data_red-sfr}).  The large pixels have an exponentially declining star formation rate with long $\tau$ compared to the ages, so the star formation rate of any given pixel does not appreciably decrease over several hundred million years.  This is not entirely unexpected: if a pixel combines several different epochs of star formation, the best fit for the combination is not representative of the overall SFH, and is likely to be young (because of the UV emission) and have a longer tau (because of the presence of older stars).
Because of this near-constant SFR for each pixel, the overall SFR accumulates over time until the present rather than showing bursty behavior.

These SFHs are the best that can be done given this paper's approach to modeling the SMC.  Siegel et~al.\ (2016, in preparation) uses the same data set to model colour-magnitude diagrams (CMDs) of individual stars in the SMC using StarFISH \citep{harris01}.  Instead of assuming a functional form of the SFH, this technique enables the SFH to be derived empirically.

In Fig.~\ref{fig-sfh_sfr} and~\ref{fig-sfh_pix}, we also plot comparisons to SFHs in the literature.  There have been a multitude of studies of the SFH of the SMC, but the large majority of them only focus on ages older than $\sim$500~Myr \citep[e.g.,][]{gardiner92,piatti12,cignoni12,weisz13,rezaeikh14} or only a small fraction of the SMC \citep[e.g.,][]{noel09,cignoni11}.  Only a handful overlap with the recent star formation that we can probe with our UV data \citep{harris04,indu11,rubele15}, though only \citet{harris04} and \citet{rubele15} derive an explicit SFH.

The recent SFHs from \citet{harris04} and \citet{rubele15} are quite similar to each other, and they are of the same order as we find.  Their SFRs are slightly larger than for our star-forming regions, though as already discussed, the SFRs derived from our star-forming regions are suppressed due to our selection techniques.
When compared to our SFH for the large pixels, the literature values have a similar slope as our measurements between 1~and 200~Myr, but beyond 200~Myr, our SFR drops more rapidly.  Overall, given that we make our SFH measurements by modeling the SEDs of broad regions, whereas \citet{harris04} and \citet{rubele15} model the CMDs of individual stars, the agreement is quite reasonable.

\citet{harris04} finds evidence for a burst of star formation 60~Myr ago, and \citet{rubele15} finds an enhanced SFR at a similar age of 40~Myr ago.  From the SFH of the star-forming regions, we also see an elevated SFR at 30-80~Myr,  We also see an additional peak at 6-10~Myr that is not apparent in the other SFHs, which can be attributed to the increased sensitivity of UV observations to younger populations.
Several studies also find a peak in the SFR at 400~Myr \citep{gardiner92,noel09,harris04}
which likely corresponds to the most recent perigalactic passage of the SMC around the Milky Way \citep{lin95}.  Our SFH for the star-forming regions has a small enhancement at 400~Myr, which may correspond to this interaction, but as discussed in \S\ref{sec-data_red-sfr}, our detection method is not very sensitive to regions this old.



\section{Summary} \label{sec-summary}

We have presented the first analysis of the SUMaC (Swift UV Survey of the Magellanic Clouds) survey, the highest resolution multi-wavelength UV survey of the Clouds yet obtained.  We have modeled the UV to NIR SEDs of star-forming regions and large $200''$ (58~pc) pixels in the SMC to extract information about the ages, masses, and shapes of the dust extinction curves.  Below are our main conclusions.

\begin{enumerate}

\item[(i)]
The 2175~\AA\ bump strength has a large-scale gradient across the face of the SMC, from weaker in the southwest to stronger in the northeast.

\item[(ii)]
The dust extinction curve is fairly steep, with $R_V < 3.1$ for the majority of the SMC, consistent with \citet{gordon03} at the overlapping locations.  There is no clear spatial trend in the $R_V$ values.

\item[(iii)]
We find evidence for elevated star formation at 6-10~Myr, 30-80~Myr (consistent with \citealt{harris04} and \citealt{rubele15}), and possibly at 400~Myr (consistent with \citealt{gardiner92}, \citealt{noel09}, and \citealt{harris04}).

\end{enumerate}


Looking to the future, there are several ways to expand upon this study.  First, we would like to enlarge the \textit{Swift}/UVOT survey to include the SMC wing and isolated star-forming regions to the east of the SMC (e.g., NGC~460 and NGC~465).  Since we find a systematically larger 2175~\AA\ bump in the northeastern SMC, it would be interesting to determine if that continues into the wing.  The star formation in NGC~460/465 and other smaller regions are in unique environments while still being a part of the SMC, and further dust curve analysis will help reveal their evolutionary history.

Second, we can include mid- and far-IR data from Spitzer \citep[SAGE-SMC;][]{gordon11} and Herschel \citep[HERITAGE;][]{meixner13} to probe the emission from dust in the SMC.  In this paper, we modeled SEDs blueward of 3.6~$\mu$m, but to fully understand the dust, one needs to include both ultraviolet absorption and infrared emission.

Finally, to fully measure the dust extinction curve, we need to acquire wide-field FUV imaging of the SMC.  GALEX completed a survey of the Magellanic Clouds \boldchange{(\citealt{simons14}; Seibert \& Schiminovich, in preparation)}, but it was after the FUV detector stopped functioning.  Astrosat's Ultraviolet Imaging Telescope \citep[UVIT;][]{hutchings14} has multi-wavelength NUV and FUV filters, and with a wide field of view ($28'$), and we strongly advocate for a survey of the Clouds.



\section*{Acknowledgements}

We thank the anonymous referee for comments that improved this paper.
We acknowledge support from NASA Astrophysics Data Analysis grant NNX12AE28G.
and sponsorship at PSU by NASA contract NAS5-00136. 
We thank Phil Massey for providing fully-reduced FITS files of his optical SMC data.
The Institute for Gravitation and the Cosmos is supported by the Eberly College of Science and the Office of the Senior Vice President for Research at the Pennsylvania State University. 
This research made use of Montage, funded by the National Aeronautics and Space Administration's Earth Science Technology Office, Computational Technologies Project, under Cooperative Agreement Number NCC5-626 between NASA and the California Institute of Technology. The code is maintained by the NASA/IPAC Infrared Science Archive.
We thank the IRSA help desk for support of the online version of Montage.
%




\bibliographystyle{mnras}

\bibliography{bibliography_file}

\bsp	
\label{lastpage}
\end{document}

%% file: table3.tex
\begin{table*}
\centering
\caption{Photometry of star-forming regions.  Magnitudes are corrected for Milky Way extinction.  An extract of the table is shown here for guidance. It is presented in its entirety in the electronic edition of the journal.}
\label{tab-sf_phot}

\scriptsize

\begin{tabular}{c ccc ccc c c}
\hline
 & \multicolumn{8}{c}{AB Magnitude} \\
\cline{2-9}
Region & $uvw2$ & $uvm2$ & $uvw1$ & B & V & R & J & 3.6~$\mu$m \\
\hline
1 & $13.053 \pm 0.050$ & $13.005 \pm 0.050$ & $13.031 \pm 0.050$ & $12.888 \pm 0.050$ & $13.058 \pm 0.050$ & $13.199 \pm 0.050$ & $13.482 \pm 0.111$ & $14.668 \pm 0.050$ \\ 
2 & $13.272 \pm 0.050$ & $13.287 \pm 0.050$ & $13.226 \pm 0.050$ & $12.324 \pm 0.050$ & $12.197 \pm 0.050$ & $12.121 \pm 0.050$ & $12.243 \pm 0.050$ & $13.711 \pm 0.050$ \\ 
3 & $12.909 \pm 0.050$ & $12.921 \pm 0.050$ & $12.776 \pm 0.050$ & $12.011 \pm 0.050$ & $12.186 \pm 0.050$ & $12.419 \pm 0.050$ & $13.095 \pm 0.058$ & $14.899 \pm 0.050$ \\ 
4 & $12.652 \pm 0.050$ & $12.667 \pm 0.050$ & $12.626 \pm 0.050$ & $11.779 \pm 0.050$ & $11.507 \pm 0.050$ & $11.286 \pm 0.050$ & $10.606 \pm 0.050$ & $11.431 \pm 0.050$ \\ 
5 & $11.551 \pm 0.050$ & $11.581 \pm 0.050$ & $11.563 \pm 0.050$ & $11.264 \pm 0.050$ & $11.255 \pm 0.050$ & $11.193 \pm 0.050$ & $11.113 \pm 0.050$ & $12.111 \pm 0.050$ \\ 
6 & $11.939 \pm 0.050$ & $11.961 \pm 0.050$ & $11.958 \pm 0.050$ & $11.886 \pm 0.050$ & $12.133 \pm 0.050$ & $12.146 \pm 0.050$ & $12.835 \pm 0.055$ & $13.373 \pm 0.050$ \\ 
7 & $10.723 \pm 0.050$ & $10.717 \pm 0.050$ & $10.731 \pm 0.050$ & $10.624 \pm 0.050$ & $10.724 \pm 0.050$ & $10.711 \pm 0.050$ & $10.992 \pm 0.050$ & $11.435 \pm 0.050$ \\ 
8 & $10.476 \pm 0.050$ & $10.468 \pm 0.050$ & $10.491 \pm 0.050$ & $9.688 \pm 0.050$ & $9.635 \pm 0.050$ & $9.689 \pm 0.050$ & $9.742 \pm 0.050$ & $11.172 \pm 0.050$ \\ 
9 & $9.326 \pm 0.050$ & $9.316 \pm 0.050$ & $9.357 \pm 0.050$ & $8.977 \pm 0.050$ & $8.836 \pm 0.050$ & $8.948 \pm 0.050$ & $8.836 \pm 0.050$ & $9.681 \pm 0.050$ \\ 
10 & $10.527 \pm 0.050$ & $10.596 \pm 0.050$ & $10.590 \pm 0.050$ & $10.517 \pm 0.050$ & $10.754 \pm 0.050$ & $10.912 \pm 0.050$ & $11.348 \pm 0.050$ & $12.407 \pm 0.050$ \\ 
\hline
\end{tabular}

\end{table*}

%% file: table4.tex
\begin{table*}
\centering
\caption{Photometry of large pixels.  Magnitudes are corrected for Milky Way extinction.  The 8~$\mu$m data is not used in the SED modeling, but is included here for reference.  An extract of the table is shown here for guidance. It is presented in its entirety in the electronic edition of the journal.}
\label{tab-pix_phot}

\scriptsize

\begin{tabular}{c ccc ccc cc}
\hline
 & \multicolumn{8}{c}{AB Magnitude} \\
\cline{2-9}
Pixel & $uvw2$ & $uvm2$ & $uvw1$ & B & V & R & 3.6~$\mu$m & 8~$\mu$m \\
\hline
1 & $13.005 \pm 0.050$ & $12.873 \pm 0.050$ & $13.039 \pm 0.050$ & $11.669 \pm 0.050$ & $11.655 \pm 0.050$ & $11.486 \pm 0.050$ & $12.370 \pm 0.050$ & $12.937 \pm 0.050$ \\ 
2 & $12.281 \pm 0.050$ & $12.266 \pm 0.050$ & $12.234 \pm 0.050$ & $10.793 \pm 0.050$ & $10.336 \pm 0.050$ & $10.011 \pm 0.050$ & $10.758 \pm 0.050$ & $11.785 \pm 0.050$ \\ 
3 & $12.828 \pm 0.050$ & $12.787 \pm 0.050$ & $12.775 \pm 0.050$ & $11.300 \pm 0.050$ & $10.987 \pm 0.050$ & $10.931 \pm 0.050$ & $12.031 \pm 0.050$ & $12.479 \pm 0.050$ \\ 
4 & $12.940 \pm 0.050$ & $12.897 \pm 0.050$ & $13.043 \pm 0.050$ & $11.870 \pm 0.050$ & $11.925 \pm 0.050$ & $11.665 \pm 0.050$ & $12.464 \pm 0.050$ & $12.778 \pm 0.050$ \\ 
5 & $12.059 \pm 0.050$ & $12.008 \pm 0.050$ & $12.044 \pm 0.050$ & $11.241 \pm 0.050$ & $11.231 \pm 0.050$ & $11.134 \pm 0.050$ & $12.142 \pm 0.050$ & $12.785 \pm 0.050$ \\ 
6 & $12.277 \pm 0.050$ & $12.281 \pm 0.050$ & $12.271 \pm 0.050$ & $11.519 \pm 0.050$ & $11.561 \pm 0.050$ & $11.460 \pm 0.050$ & $12.007 \pm 0.050$ & $12.639 \pm 0.050$ \\ 
7 & $12.891 \pm 0.050$ & $12.855 \pm 0.050$ & $12.858 \pm 0.050$ & $11.712 \pm 0.050$ & $11.713 \pm 0.050$ & $11.498 \pm 0.050$ & $11.935 \pm 0.050$ & $12.356 \pm 0.050$ \\ 
8 & $13.077 \pm 0.050$ & $12.983 \pm 0.050$ & $13.041 \pm 0.050$ & $11.276 \pm 0.050$ & $10.853 \pm 0.050$ & $10.584 \pm 0.050$ & $11.442 \pm 0.050$ & $12.164 \pm 0.050$ \\ 
9 & $11.189 \pm 0.050$ & $11.205 \pm 0.050$ & $11.202 \pm 0.050$ & $10.598 \pm 0.050$ & $10.610 \pm 0.050$ & $10.575 \pm 0.050$ & $11.838 \pm 0.050$ & $12.469 \pm 0.050$ \\ 
10 & $11.257 \pm 0.050$ & $11.282 \pm 0.050$ & $11.238 \pm 0.050$ & $10.778 \pm 0.050$ & $10.850 \pm 0.050$ & $10.815 \pm 0.050$ & $12.045 \pm 0.050$ & $12.133 \pm 0.050$ \\ 
\hline
\end{tabular}

\end{table*}

%% file: table5.tex
\begin{table*}
\centering
\caption{Physical properties of star-forming regions.  An extract of the table is shown here for guidance. It is presented in its entirety in the electronic edition of the journal.}
\label{tab-sf_prop}

\begin{tabular}{l l cccccc}
\hline
Region & Radius & $A_V$ & $R_V$ & Bump & Log Age & Log Mass & Log Stellar Mass \\
  & (pc) & (mag) &  &  & (Myr) & (\msun) & (\msun) \\
\hline
1 & 9.0 & 0.072$^{+0.028} _{-0.021}$ & 2.023$^{+0.722} _{-0.381}$ & 0.571$^{+0.701} _{-0.407}$ & 1.118$^{+0.029} _{-0.028}$ & 2.868$^{+0.049} _{-0.044}$ & 2.81$^{+0.05} _{-0.04}$ \\ 
2 & 9.3 & 0.088$^{+0.019} _{-0.015}$ & 1.694$^{+0.289} _{-0.143}$ & 1.117$^{+0.535} _{-0.510}$ & 1.839$^{+0.007} _{-0.015}$ & 3.678$^{+0.012} _{-0.014}$ & 3.58$^{+0.01} _{-0.01}$ \\ 
3 & 9.6 & 0.195$^{+0.020} _{-0.018}$ & 1.564$^{+0.098} _{-0.048}$ & 1.077$^{+0.295} _{-0.249}$ & 0.777$^{+0.002} _{-0.004}$ & 2.775$^{+0.013} _{-0.013}$ & 2.75$^{+0.01} _{-0.01}$ \\ 
4 & 10.9 & 1.144$^{+0.099} _{-0.131}$ & 4.434$^{+0.450} _{-0.595}$ & 0.143$^{+0.122} _{-0.092}$ & 1.729$^{+0.092} _{-0.124}$ & 4.355$^{+0.096} _{-0.138}$ & 4.26$^{+0.10} _{-0.14}$ \\ 
5 & 17.3 & 0.351$^{+0.168} _{-0.113}$ & 2.981$^{+0.766} _{-0.723}$ & 0.366$^{+0.289} _{-0.212}$ & 1.010$^{+0.059} _{-0.039}$ & 3.540$^{+0.164} _{-0.034}$ & 3.49$^{+0.16} _{-0.03}$ \\ 
6 & 14.0 & 0.948$^{+0.061} _{-0.064}$ & 5.292$^{+0.146} _{-0.220}$ & 0.168$^{+0.142} _{-0.109}$ & 0.697$^{+0.007} _{-0.009}$ & 3.159$^{+0.026} _{-0.036}$ & 3.14$^{+0.03} _{-0.04}$ \\ 
7 & 22.9 & 0.259$^{+0.132} _{-0.162}$ & 4.035$^{+0.749} _{-0.802}$ & 0.453$^{+0.586} _{-0.311}$ & 1.226$^{+0.053} _{-0.049}$ & 4.007$^{+0.031} _{-0.038}$ & 3.95$^{+0.03} _{-0.04}$ \\ 
8 & 25.1 & 0.332$^{+0.053} _{-0.040}$ & 1.805$^{+0.255} _{-0.187}$ & 0.150$^{+0.139} _{-0.098}$ & 0.890$^{+0.015} _{-0.026}$ & 3.975$^{+0.018} _{-0.020}$ & 3.94$^{+0.02} _{-0.02}$ \\ 
9 & 40.1 & 0.499$^{+0.152} _{-0.167}$ & 3.103$^{+0.748} _{-0.664}$ & 0.142$^{+0.162} _{-0.099}$ & 1.020$^{+0.068} _{-0.067}$ & 4.507$^{+0.197} _{-0.049}$ & 4.46$^{+0.20} _{-0.05}$ \\ 
10 & 26.4 & 0.026$^{+0.065} _{-0.020}$ & 4.764$^{+0.567} _{-1.330}$ & 0.639$^{+0.804} _{-0.481}$ & 0.823$^{+0.005} _{-0.008}$ & 3.395$^{+0.022} _{-0.011}$ & 3.36$^{+0.02} _{-0.01}$ \\ 
\hline
\end{tabular}

\end{table*}

%% file: table6.tex
\begin{table*}
\centering
\caption{Physical properties of large pixels.  An extract of the table is shown here for guidance. It is presented in its entirety in the electronic edition of the journal.}
\label{tab-pix_prop}

\begin{tabular}{l ccccccc}
\hline
Pixel & $A_V$ & $R_V$ & Bump & Log Age & $\tau$ & Log Mass & Log Stellar Mass \\
  & (mag) &  &  & (Myr) & (Myr) & (\msun) & (\msun) \\
\hline
1 & 0.122$^{+0.139} _{-0.094}$ & 2.302$^{+0.708} _{-0.506}$ & 0.139$^{+0.719} _{-0.110}$ & 3.468$^{+0.228} _{-0.238}$ & 1960$^{+1430} _{-1250}$ &  4.40$^{+ 1.14} _{- 0.20}$ & 4.36$^{+1.14} _{-0.20
}$ \\ 
2 & 0.249$^{+0.188} _{-0.102}$ & 2.219$^{+0.489} _{-0.442}$ & 0.208$^{+0.330} _{-0.151}$ & 3.904$^{+0.156} _{-0.291}$ & 2110$^{+1260} _{-860}$ &  5.30$^{+ 1.00} _{- 0.15}$ & 5.24$^{+1.00} _{-0.15
}$ \\ 
3 & 0.287$^{+0.046} _{-0.054}$ & 1.598$^{+0.141} _{-0.073}$ & 0.183$^{+0.198} _{-0.121}$ & 3.133$^{+0.131} _{-0.099}$ & 1920$^{+1470} _{-1150}$ &  4.43$^{+ 1.11} _{- 0.23}$ & 4.40$^{+1.11} _{-0.23
}$ \\ 
4 & 0.165$^{+0.243} _{-0.140}$ & 4.024$^{+1.001} _{-0.865}$ & 0.229$^{+0.717} _{-0.179}$ & 3.413$^{+0.142} _{-0.284}$ & 1980$^{+1430} _{-1110}$ &  4.31$^{+ 1.22} _{- 0.21}$ & 4.27$^{+1.22} _{-0.20
}$ \\ 
5 & 0.141$^{+0.143} _{-0.101}$ & 2.948$^{+1.013} _{-0.668}$ & 0.293$^{+0.693} _{-0.223}$ & 2.925$^{+0.185} _{-0.303}$ & 2010$^{+1360} _{-1250}$ &  4.23$^{+ 1.04} _{- 0.28}$ & 4.21$^{+1.04} _{-0.28
}$ \\ 
6 & 0.670$^{+0.105} _{-0.098}$ & 4.887$^{+0.395} _{-0.415}$ & 0.218$^{+0.215} _{-0.147}$ & 2.644$^{+0.134} _{-0.163}$ & 1990$^{+1380} _{-1280}$ &  4.14$^{+ 1.07} _{- 0.29}$ & 4.12$^{+1.07} _{-0.29
}$ \\ 
7 & 0.757$^{+0.124} _{-0.154}$ & 4.064$^{+0.489} _{-0.370}$ & 0.138$^{+0.171} _{-0.098}$ & 3.023$^{+0.216} _{-0.237}$ & 1880$^{+1520} _{-1150}$ &  4.35$^{+ 1.19} _{- 0.26}$ & 4.32$^{+1.19} _{-0.26
}$ \\ 
8 & 0.283$^{+0.117} _{-0.076}$ & 1.727$^{+0.340} _{-0.163}$ & 0.091$^{+0.194} _{-0.069}$ & 3.837$^{+0.165} _{-0.227}$ & 1990$^{+1380} _{-810}$ &  5.02$^{+ 1.03} _{- 0.13}$ & 4.96$^{+1.03} _{-0.13
}$ \\ 
9 & 0.092$^{+0.095} _{-0.058}$ & 2.612$^{+0.957} _{-0.718}$ & 0.786$^{+0.713} _{-0.509}$ & 2.487$^{+0.208} _{-0.203}$ & 2080$^{+1160} _{-1230}$ &  4.16$^{+ 0.75} _{- 0.29}$ & 4.15$^{+0.75} _{-0.29
}$ \\ 
10 & 0.079$^{+0.080} _{-0.050}$ & 3.449$^{+1.101} _{-1.066}$ & 1.003$^{+0.640} _{-0.611}$ & 2.406$^{+0.124} _{-0.135}$ & 2090$^{+1190} _{-1290}$ &  4.02$^{+ 0.79} _{- 0.29}$ & 4.01$^{+0.79} _{-0.29
}$ \\ 
\hline
\end{tabular}

\end{table*}